\documentclass[a4paper,11pt]{article}
\usepackage{jheppub}
\usepackage{graphicx}
\usepackage{amsmath}
\usepackage{amsfonts}
\usepackage{amssymb}
\usepackage{slashed}
\usepackage[colorlinks=true,linkcolor=blue]{hyperref}
\usepackage{showlabels}
\usepackage{bm}
\usepackage{empheq}
\usepackage[super]{nth}
\def\sss{\scriptscriptstyle}

\def\tA{\theta_{\!\sss{A}}}
\def\tAext{\theta_{\!\sss{A},\mathrm{ext}}}

\def\rmax{r_{\mathrm{max}}}
\def\kmin{\kappa_{\mathrm{min}}}
\def\kmax{\kappa_{\mathrm{max}}}
\def\barz{{\bar{z}_2}}

\def\OO{\mathcal{O}}
\def\LL{\mathcal{L}}
\def\be{\begin{equation}}
\def\ee{\end{equation}}
\def\bea{\begin{eqnarray}}
\def\eea{\end{eqnarray}}

\def\Eq#1{Eq.~\eqref{#1}}

\def\atan{\mathrm{atan}}

\title{Intercommutation of U(1) global cosmic strings}
\author{Guy D.\ Moore}

\affiliation{Institut f\"ur Kernphysik, Technische Universit\"at Darmstadt\\
Schlossgartenstra{\ss}e 2, D-64289 Darmstadt, Germany}
\emailAdd{guy.moore@physik.tu-darmstadt.de}

\abstract{
Global strings (those which couple to Goldstone modes) may play a role
in cosmology.  In particular, if the QCD axion exists, axionic strings
may control the efficiency of axionic dark matter abundance.
The string network dynamics depend on the string intercommutation
efficiency (whether strings re-connect when they cross).  We point out
that the velocity and angle in a collision between global strings
``renormalize'' between the network scale and the microscopic scale,
and that this plays a significant role in their intercommutation
dynamics.  We also point out a subtlety in treating intercommutation
of very nearly antiparallel strings numerically.  We find that the
global strings of a O(2)-breaking scalar theory do intercommute for
all physically relevant angles and velocities.
}

\keywords{axions, dark matter, cosmic strings, global strings}
\begin{document}
\maketitle
\section{Introduction}
\label{sec:intro}

Cosmic strings \cite{Kibble:1976sj,Gibbons:1990gp} are hypothetical extended
solitonic excitations which may play a significant role in cosmology.
Their original motivation, for structure formation
\cite{Zeldovich:1980gh,Vilenkin:1981iu,Kibble:1980mv}, appears in
conflict with modern microwave sky data \cite{Ade:2013xla}.  But cosmic
strings may be important in other contexts.  In particular, if the QCD
axion \cite{Weinberg:1977ma,Wilczek:1977pj} exists, the axion field
may contain a string network in the early Universe \cite{Davis:1986xc}
which may dominate axion production and play a central role in the
axion as a dark matter candidate.

String defects can occur whenever the vacuum spontaneously breaks a
symmetry, say, breaking $G$ down to $H\in G$, such that the quotient
group (vacuum manifold) $G/H$ has nontrivial $\pi_1$ homotopy.  The
simplest example is the complete breaking of an SO(2) or U(1)
symmetry. Generally, strings are divided into two sorts; ``local''
strings, which typically occur when $G$ is a gauge group and which do
not couple to any massless fields \cite{Kibble:1976sj}, and ``global''
strings, which typically occur when $G$ is a global (non-gauged)
symmetry, in which case the strings couple to the associated massless
Goldstone bosons \cite{Vilenkin:1982ks}.  There is a rich literature
studying local strings and their networks
\cite{Bennett:1989ak,Allen:1990tv,Vanchurin:2005yb}.  This includes a
rather careful study of when crossing strings intercommute with each
other \cite{Laguna:1990it,Bettencourt:1996qe,Achucarro:2006es,
  Achucarro:2010ub}.

\begin{figure}[htb]
  \centerline{\epsfxsize=0.5\textwidth\epsfbox{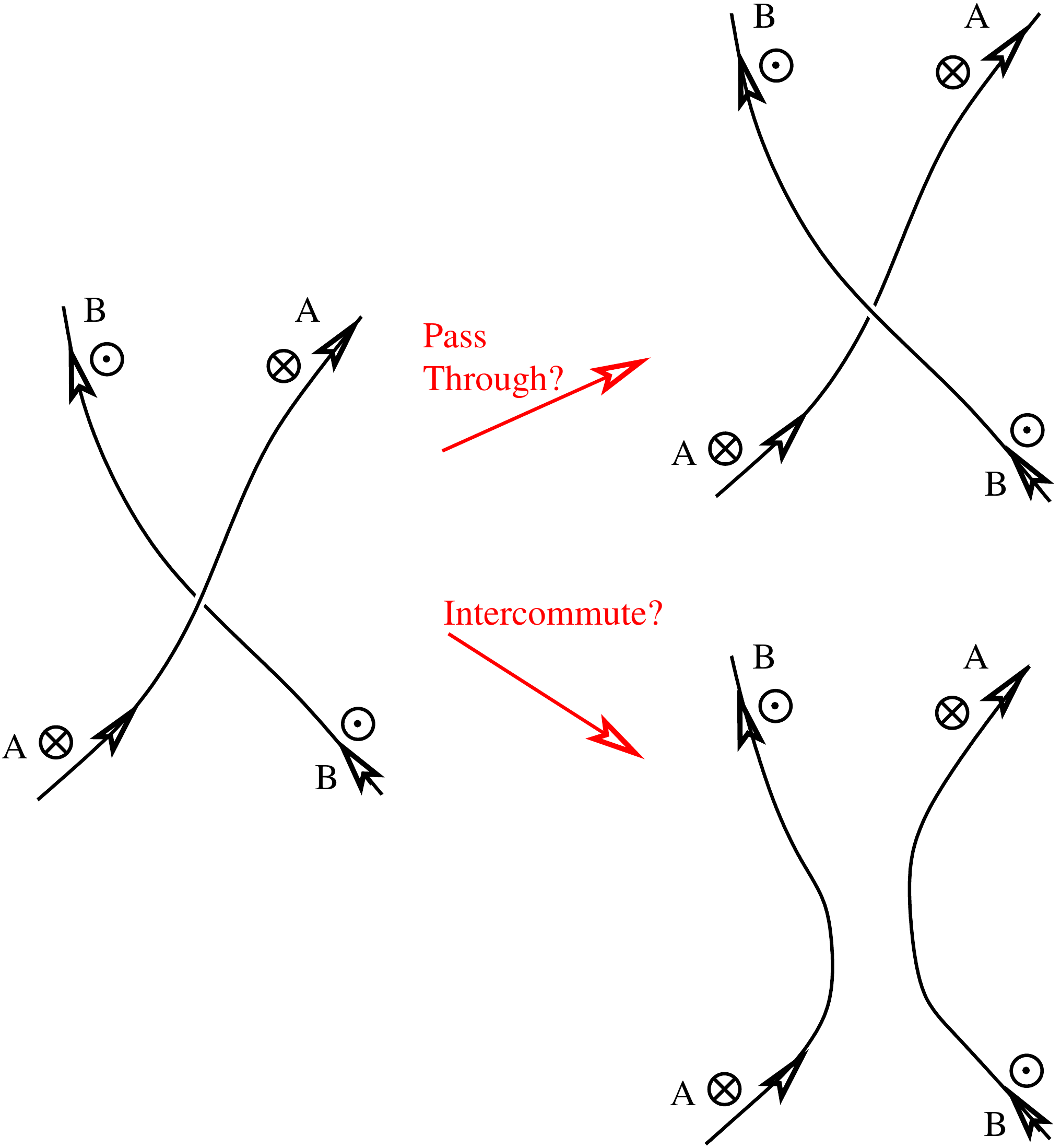}}
  \caption{\label{fig:intercommute}
    Left: two strings approach, with $A$ descending and $B$ rising out
    of the page.  The arrows indicate the ``sense'' of the string.
    There are two possibilities:  they pass through each other (top
    right), or they intercommute (bottom right).  Which occurs is
    determined by the microphysics at the point where they touch.}
\end{figure}

Global strings, such as the axionic string networks alluded to above,
are harder to study.  The strings interact with each
other via massless fields, which change the network's dynamics, both
by allowing strings to radiate energy efficiently
\cite{Vilenkin:1986ku,Battye:1993jv,Battye:1995hw,Hagmann:2000ja}, and by
communicating inter-string forces.
To simulate such a network faithfully requires a hybrid algorithm,
which treats the string cores via a Nambu-Goto action and treats the
Goldstone field with lattice methods \cite{Dabholkar:1989ju}, together with
some string-field interaction.  We have recently presented such an
algorithm \cite{axion2}.  However, to implement it we must know when
strings which cross each other will inter-commute, and when they will
simply pass through each other (see Figure \ref{fig:intercommute}).
This problem has been previously addressed via microscopic simulations
\cite{Shellard:1987bv}.  But we will argue here that the physics is more
complicated due to the long range inter-string interactions, and the
issue of string inter-commutation deserves some more study.

\begin{figure}[htb]
  \centerline{\epsfxsize=0.9\textwidth\epsfbox{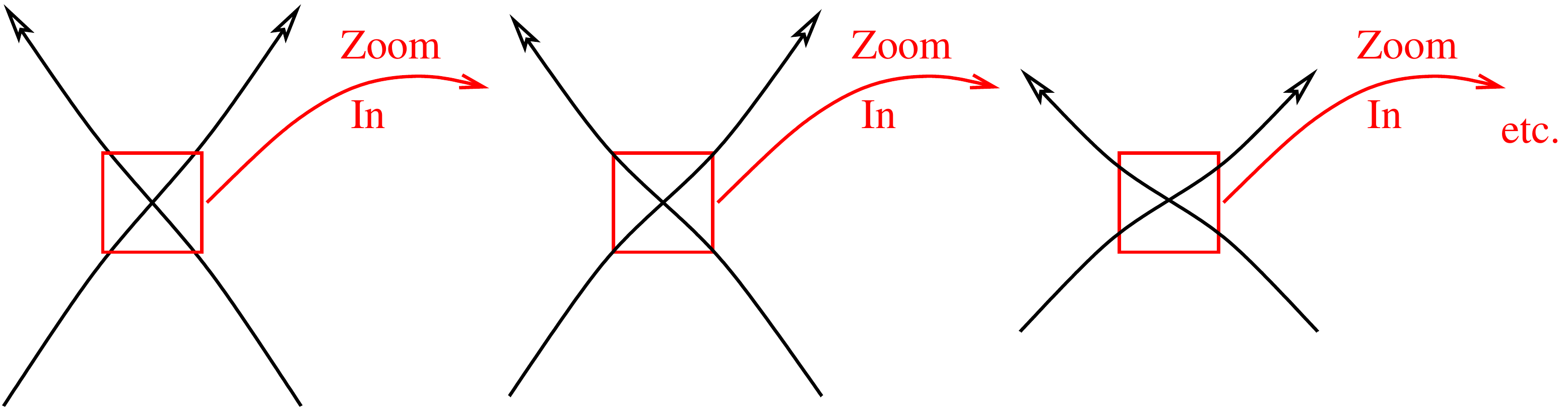}}
  \caption{\label{fig:scales}
    Two string just before crossing.  They twist towards larger
    crossing angle; zooming in, we see that at short distances they
    twist more; zooming in, they twist still more, and so on.  Even if
    the twist is slight, over many orders of magnitude in scale it can
    become very large.}
\end{figure}

Cosmic string networks involve two disparate scales.  There is a
microscopic scale $m_h^{-1}$, set by the inverse mass of the (Higgs) radial
excitation of the symmetry-breaking field.  And there is a macroscopic
scale, the average inter-string separation $L$, which is typically of
order the Hubble scale $H^{-1}$.  For axions around the time where the
dark matter density is established, these scales differ by a factor of
$\sim 10^{30}$ \cite{Sikivie:2006ni}.  The energy in a local cosmic string is
carried within a few $m_h^{-1}$ of the string's core.  But for a global
string, the energy is distributed logarithmically over all scales
between $m_h^{-1}$ and $L$.  As global strings approach each other
through these intermediate scales, this energy distribution is
responsible for inter-string interactions, which apply both torques,
and attraction or repulsion, to the strings.  The angle and velocity
with which the strings approach therefore evolves logarithmically with
scale.  We illustrate this idea in Figure \ref{fig:scales}.
Here we study this evolution, and its impact on the physics of
string intercommutation.  We concentrate on strings arising from a
complex scalar (the O(2) model or relativistic 3D xy model); the
details and results could be quite different for strings with other
more complicated symmetry breaking patterns.

In the next section we review the physics of global cosmic strings,
and we explain why long range inter-string interactions can be
important.  Section \ref{sec:renorm} treats the inter-string forces
explicitly and finds how the angle and velocity of string approach
changes with scale.  This is combined with a new study of microscopic
string intercommutation in Section \ref{sec:micro}.

We end with a discussion and conclusion, but we will give the main
findings here.  At the macroscopic scale strings may approach each
other at a range of angles and relative velocities.  But accounting
for inter-string forces, at the microscopic scale of $100/m$, the
strings are essentially always almost antiparallel (the attractive
channel) and approaching with a highly relativistic velocity
$v \simeq 0.9$.  We show in contradiction to previous results
\cite{Shellard:1987bv} that such strings always intercommute.  Therefore
in treating O(2)-model global strings, one should assume that
intercommutation always occurs.

\section{Global string review}
\label{sec:review}

Here we review global strings (see also
\cite{Vilenkin:1982ks,Gibbons:1990gp,vilenkin2000}).
Consider a complex scalar field $\varphi$ with U(1) symmetry (or
equivalently, two real scalars $\varphi_r$, $\varphi_i$ with an O(2)
symmetry; $\varphi = (\varphi_r + i\varphi_i)/\sqrt{2}$).  The
most general renormalizable Lagrangian is%
\footnote{We use the $[{-},{+},{+},{+}]$ or mostly-positive metric}
\be
\label{Lagrangian}
-\LL = \partial_\mu \varphi^* \partial^\mu \varphi
 + \frac{\lambda}{8} \left( 2 \varphi^* \varphi - f_a^2 \right)^2 \,.
\ee
If $f_a^2 > 0$ then the classical vacuum
$\sqrt{2} \varphi = f_a e^{i\tA}$ spontaneously breaks the U(1)
symmetry.  The excitation spectrum about this vacuum contains a radial
``Higgs'' excitation with mass $m_h^2 = \lambda f_a^2$ and angular
excitations, which are massless Goldstone modes.  We are interested in
solitonic solutions corresponding to an extremely large number of
quanta, so a classical description of the field and its dynamics are
sufficient.

Now $\tA$ is only defined modulo $2\pi$, and there are configurations
where it cannot be defined in a continuous and single-valued way.
Specifically, a topological string solution is where there is a
1-dimensional oriented (linelike) locus of points, called a cosmic
string, where $|\varphi|=0$, and such that $\tA$ varies by $2\pi$ as
one circles the string with positive sense.  Locally the string is
nearly straight and we can use it as the $z$-axis in polar
$(z,r,\phi)$ coordinates.  The string solution is
$\varphi(z,r,\phi) = f_a h(r) \exp(i(\phi-\phi_0))$, with $h(r)$
chosen to minimize the energy per unit length (string tension)
\bea
\label{E_h}
T & = & \int r\,dr\,d\phi \left( |\nabla \varphi|^2
+ V(\varphi^* \varphi) \right)
\nonumber \\
& = & \pi f_a^2 \int r \, dr \left(
(\partial_r h)^2 + \frac{h^2}{r^2} + \frac{m_h^2}{8} (h^2-1)^2 \right) \,,
\eea
which is minimized when $h(r)$ obeys
\be
\label{h_of_r}
\partial_r^2 h + \frac{1}{r} \partial_r h + \frac{h^2}{r^2}
- \frac{m_h^2}{2} h (h^2-1) = 0 \,, \qquad
h(0)=0,\;\; \lim_{r\to \infty}h(r)=1 \,.
\ee
The most important term in \Eq{E_h} is the
$\pi f_a^2 \int r\, dr \: h^2/r^2$ term, arising from
$|\nabla_\phi \varphi|^2$.  All other terms only contribute
appreciably for $m_h r\sim 1$ and fall off quickly at large $r$.  But
this term receives equal contributions from all logarithmic scales
larger than $m_h^{-1}$:
\be
T \simeq \int_{m_h^{-1}}^{\rmax} r\,dr \: \pi f_a^2 (1/r^2)
= \pi f_a^2 \Big( \ln(m_h r_{\mathrm{max}}) + \OO(1) \Big) \,,
\label{Tension}
\ee
where $\rmax$ is a long-distance cutoff, which would physically be
provided by the curvature scale of the string or the distance to the
next string.  This logarithmic energy scaling will be essential to our
arguments.

The other essential feature of the strings is the way they interact
with their environment.  Suppose that a string exists in an
environment where $\tA$ also varies uniformly, for instance because of
the far field of another string.  Outside the string core the solution
is described by $\tA$ alone, and its equation of motion is linear so
we can superpose solutions.  In this case our \textsl{Ansatz} for
the field becomes
\be
\label{Ansatz}
\varphi(x) = f_a h(r) e^{i\phi} e^{i\tAext} \,, \quad
\tAext \simeq x_i \nabla_i \tAext \,,
\ee
where we have approximated the external field by the first term in its
Taylor series.  Near the string core the equation of motion is not
linear, so $\nabla_i \tAext$ will cause an acceleration in the
string.  The easy way to determine this is to find the force per unit
length on the string, by integrating the stress normal to a boundary
around the string which we draw at some radius $r \gg m_h^{-1}$:
\be
\label{dF}
dF_i = \int r d\phi \; T_{ij}(r,\phi) \, \hat{n}_j \,.
\ee
Here $T_{ij}$ is the stress tensor
\be
\label{Tij}
T_{ij} = f_a^2 \left( \partial_i \tA \partial_j \tA
- \frac{1}{2} \delta_{ij} \partial_k \tA \partial_k \tA \right) \,.
\ee
Now $\partial_i \tA = \partial_i \tAext + \hat\phi_i / r$.  The
$(\tAext)^2$ term and the $(\hat\phi)^2$ term each integrate to zero,
but the cross term contributes
\be
\label{dFis1}
dF_i = 2\pi f_a^2 \epsilon_{ijz} \nabla_j \tAext \,.
\ee
Now $z$ appears because it is the unit tangent of the string; so this
generalizes for a string of unit tangent $\hat{s}$ to
\be
\label{dFis2}
dF_i = 2\pi f_a^2 \epsilon_{ijk} \nabla_j \tAext s_k \,,
\ee
or for a moving string with 4-velocity $v^\mu = (1,\vec{v})$, to
\be
\label{dFis3}
dF_\mu = 2\pi f_a^2 \epsilon_{\mu\nu\alpha\beta}
\nabla^\nu \tAext s^\alpha v^\beta \,.
\ee

These are the facts about strings that we will need in what follows.

\section{Renormalization of string angle and velocity}
\label{sec:renorm}

Consider two strings, approaching each other and at some relative
angle.  When the string separation $z$ is small compared to the
characteristic inter-string spacing $L$, the strings can be taken as
nearly straight and moving with nearly uniform velocity.  Then we can
work in the frame where one string stretches in the $y$ direction and
the other in the $xy$ plane, and they approach each other along the
$z$ direction with equal and opposite velocities $v$.  The system is
fully specified by the velocity $v$, the separation $z$ (or
equivalently the time to impact $t=z/2v$ with the 2 because each
string is moving), and the angle $\phi$, describing how far from 
parallel the two strings are.

The strings exert forces on each other.  The $z$ component of the
force is repulsive if $\phi < \pi/2$ (nearly parallel) and attractive
if $\phi > \pi/2$ (nearly antiparallel).  There are also forces in the
$xy$ plane, which tend to push the strings into the antiparallel
relative orientation.  We want to understand how these forces change
the relative velocity and orientation of the strings, as the
inter-string distance drops from the macroscopic to the microscopic
scale.

The characteristic size of this force is
$dF \sim 2\pi f_a^2 / z$, while the tension of the string, which sets
its inertia, is $T = \pi f_a^2 \ln(zm_h)$.  Therefore, in the time
$t\sim z/2v$ which it takes for the strings to get a factor of 2
closer together, the velocity and orientation of the strings can
change by an amount $t\, dF/T \sim 1/[v \ln(zm_h)]$.  This is small by
one power of our large logarithm.  But the number of factors-of-two
over which the strings must approach each other is large,
$\sim \ln(zm_h)$.  Therefore, even if a factor-of-2 change in the
separation only makes a small correction to the relative velocity and
angle near the intersection point, the strings' relative velocity and
angle of approach will change significantly as the separation goes
from the macroscopic to the microscopic scale.  This is analogous to
renormalization group flow.  At each scale, it is the $v,\phi$ value
at that scale which is relevant.  Over a factor of 2 change in scale,
these change by a small amount, but the cumulative change can be large
if the log of the ratio of scales is large enough.  In this section we
will find a differential equation for how $v,\phi$ evolve with the log
of the separation scale.

First we estimate of whether the change will be large.
Introducing $\kappa = \ln(zm_h)$, the total change in velocity and
angle is of order
\be
\label{handwave}
\Delta (v,\phi) \sim
\int_{\kmin}^{\kmax}
\frac{d\kappa}{\kappa} \sim \ln \frac{\kmax}
     {\kmin} \,.
\ee
For axionic strings at a temperature around 1GeV we have
$\kmax \sim \ln(f_a/H) \sim \ln(10^{30})\sim 70$.
The short distance physics can be studied with lattice methods, but
only using
$\kmin \sim \ln(N_{\mathrm{sites}}) \sim \ln(10^2) \sim 5$.  Then
$\ln(\kmax/\kmin) \simeq \ln(70/5) \simeq 2.6$ is actually fairly large.

Let us move forward to a quantitative calculation of how the strings'
relative angle and velocity change with scale.  We will assume that
the string is \textsl{nearly} straight, and treat both the string's
curvature and the force on the string to be $\OO(\kappa^{-1})$.  We
work to order $\kappa^{-1}$, which means we may treat the string as
straight and its motion as uniform in computing the force per unit
length on each string.  First we compute this force.  For the string
on the $y$ axis at the moment when its $z$-coordinate is $z_0$, in its
rest frame we have $\tA = \atan(x/(z-z_0))$.  So in the frame where it
moves in the $+z$ direction with velocity $v$ and at the moment $t_0$
when the string is at $z_0$, we have
\be
\label{gradients}
\tA = \atan \frac{x}{\gamma([z{-}z_0]-v[t{-}t_0])} \,, \quad
\partial_x \tA = \frac{\gamma [z{-}z_0]}{x^2 + \gamma^2 [z{-}z_0]^2} \,, \quad
\partial_z \tA = \frac{-\gamma x}{x^2 + \gamma^2 [z{-}z_0]^2} \,, \qquad
\partial_t \tA = -v \partial_z \tA \,.
\ee
Here $\gamma = 1/\sqrt{1-v^2}$ as usual.

We take the point in the $(x,y)$ plane where the strings will cross to
be $(x,y)=(0,0)$.  In the approximation that the upper string is
straight, the position varies with the length $\ell$ along the string
from this point as: $z-z_0\equiv z =2vt$ and
$(x,y) = \ell (\sin\phi,\cos\phi)$.  At this point, the force on the
upper string, using \Eq{dFis3} and \Eq{gradients}, is
\begin{eqnarray}
\label{Fresult}
dF_\mu & = & 2\pi f_a^2 \epsilon_{\mu\nu\alpha\beta} v^\nu \nabla^\alpha \tA
s^\beta \,, \nonumber \\
dF_\perp & = & 2\pi f_a^2 \frac{\gamma (1+v^2) \ell \sin\phi}
{\gamma^2 z^2 + \ell^2 \sin^2 \phi} \,, \nonumber \\
dF_z & = & 2\pi f_a^2 \frac{\gamma z \cos\phi}
{\gamma^2 z^2 + \ell^2 \sin^2 \phi} \,, \nonumber \\
dF_0 & = & -2\pi f_a^2 \frac{\gamma v z \cos\phi}
{\gamma^2 z^2 + \ell^2 \sin^2 \phi} \,.
\end{eqnarray}
Here $dF_0$ is the energy exchange rate, and the relative sign between
$dF_z$ and $dF_0$ is because the upper string moves in the $-z$
direction.

Beyond lowest order, the string is not straight.  Write the string
separation a time $t$ before the strings meet as
\be
\label{zoft}
z(\ell,t) = z_1(t) + t \barz(\ell/t) \,,
\ee
where $\partial_t z_1 = -2v$ with $v$ the velocity at the crossing point.
Here $\barz$ parameterizes how the velocity varies along the string,
which should be a function of $\ell/t$ only as the appearance should
be self-similar as we vary scale.  Both $\barz$ 
and $t\partial_t v$ are first-order small.
Similarly, rotating the $(x,y)$ plane so the strings are at angles of
$\pm \phi/2$ with respect to the $y$ axis, we have
\be
\label{phioft}
x(\ell,t) = \int_0^\ell d\ell' \sin(\phi(\ell',t)/2) \,,
\qquad
\phi(\ell,t) = \phi_1(t) + \phi_2(\ell/t) \,.
\ee
Note that we are defining $\phi(\ell,t)$ in terms of the unit tangent
of the string.
We assume that $t\partial_t \phi_1$ and $\phi_2$ are first-order
small, and $\phi_2$ is a self-similar function of $\ell/t$ only.
We also introduce the notation $\xi = \ell/t$.  We will write
$\barz' = \partial_\xi \barz(\xi)$, so
$\partial_\ell \barz = t^{-1} \barz'$ and
$\partial_t \barz = -\xi t^{-1} \barz'$, and similarly for $\phi_2(\xi)$.

Introducing the energy per unit length of the string,
\be
\label{defeps}
\varepsilon = \pi \kappa f_a^2 \gamma  = \gamma T \,,
\ee
the equation of motion for the string is \cite{Turok:1984db}
\be
\label{stringEOM}
\partial_t^2 x_i = (1-v^2) \partial_{\ell}^2 x_i
+ \varepsilon^{-1} ( F_i - F_0 \partial_t x_i ) \,.
\ee
Using that $F_0 = v_i F_i$, the last term can be rewritten as
$\varepsilon^{-1} ( \delta_{ij} - v_i v_j) F_j$, where
$\varepsilon^{-1}$ accounts for the string's inertia per unit length,
and $(\delta_{ij} - v_i v_j)$ is the usual relativistic reduction of the
acceleration for a force acting along the current direction of
propagation.

Consider first the $z$-motion.  We have
\bea
\label{zmotion1}
\partial_t z &=& -2v + \barz - \xi \barz' \,, \nonumber \\
\partial_t^2 z & = & -2 \partial_t v + \frac{\xi^2}{t} \barz'' \,,
\nonumber \\
\partial_\ell z & = &  \barz' \,, \nonumber \\
\partial_\ell^2 z & = & \frac{1}{t} \barz'' \,.
\eea
Inserting these into \Eq{stringEOM} and using \Eq{Fresult}, we find
\be
\label{zmotion2}
- \partial_t v + \frac{\xi^2 - (1-v^2)}{2t} \barz''
= \frac{1-v^2}{\pi \kappa f_a^2 \gamma} F_z
= \frac{2}{\kappa} \frac{(1-v^2) z \cos\phi}
{z^2/(1-v^2) + \ell^2 \sin^2 \phi} \,.
\ee
The factor of 2 difference between \Eq{zmotion1} and the left-hand side
of \Eq{zmotion2} is because each string feels a force, or equivalently
because each string must only move a distance $z/2$ before they meet,
so \Eq{stringEOM} should be applied to $z/2$ not $z$.

This expression is a complicated differential equation for both $v(t)$
and for $\barz(\xi)$.  However, at the special point $\xi^2=(1-v^2)$,
the $\barz''$ term drops out and we directly find $\partial_t v$.
Evaluating at this point, and using $z = 2vt$ on the right hand side,
we find
\be
\label{zmotion3}
\frac{t\, dv}{\: dt} = -\frac{2}{\kappa}\: \frac{2(1-v^2)^2 v \cos\phi}
     {4v^2 + (1-v^2)^2 \sin^2 \phi} \,.
\ee
The relative velocity rises if $\phi>\pi/2$ and it falls if
$\phi<\pi/2$, as expected.

The reason that the $\barz$ terms drop out at $\xi^2=1-v^2$
is because waves propagate along the string with
velocity 1, which in the $xy$ plane is velocity $\sqrt{1-v^2}$.  The
point $\xi^2=1-v^2$ is the point where a wave will reach $\ell=0$ at
the moment the strings meet.  Larger $\xi$ values cannot communicate
with the intersection point, and any dynamics at smaller $\xi$ is
carried, as a wave, past the intersection point before the strings
collide.

Now we repeat the analysis for the in-plane (angular) evolution.
Starting with \Eq{phioft}, we find
\bea
\label{transmotion1}
\frac{d^2 x}{d\ell^2} & = & \frac{\cos(\phi/2)}{2t} \phi_2' \,,
\nonumber \\
\frac{d^2 x}{dt^2} & = &
\frac{\xi t \cos(\phi/2)}{2} \partial_t^2 \phi_1 +
\frac{\xi^2 \cos(\phi/2)}{2t} \phi_2' \,.
\eea
The expressions for $y$ are the same but with $\sin(\phi/2)$.
Therefore the transverse motion of the string is
\be
\label{transmotion2}
\frac{\xi^2 - (1-v^2)}{2t} \phi_2'
+ \frac{\xi t}{2} \partial_t^2 \phi_1 =
\frac{2}{\kappa} {\frac{(1+v^2) \ell \sin\phi}
  {z^2/(1-v^2) + \ell^2 \sin^2 \phi} } \,.
\ee
The factor of $1/2$ appearing in each term on the left-hand side arises
because each string changes orientation.  Again this is a differential
equation for both $\phi_1(t)$ and for $\phi_2(\xi)$, but $\phi_2$
drops out at the special point $\xi^2 = 1-v^2$ for the same physical
reason as before.  Evaluating at this point, we find
\be
\label{transmotion3}
t^2 \partial_t^2 \phi_1 = \frac{2}{\kappa} \:
{\frac{2(1+v^2)(1-v^2) \sin\phi}
  {4 v^2 + (1-v^2)^2 \sin^2 \phi} } \,.
\ee
The most general solution is
$\phi_1 = A \ln(t_0/t)+B + C t$, where $A$ is the right
hand side.  The term $C$ rapidly becomes subdominant as we go to small
times -- it is ``renormalization group irrelevant'' -- and we should
ignore it.  Then $B = \phi(t_0)$, and the coefficient $A$ tells how
the angle scales logarithmically with time -- or equivalently with
scale.  The angle between the strings increases for any
value of $\phi < \pi$, so the strings always become more
antiparallel.  This is as expected; objects typically rotate to be in
the attractive channel.

\Eq{zmotion3} and \Eq{transmotion3} together give us our
``renormalization group'' equations for understanding how the relative
velocity and angle of the strings change with scale.  Note that the
denominator becomes singular if $v\to 0$ and $\sin(\phi) \to 0$ at
the same time.  This is because slow-moving strings are close together
and therefore exert larger forces on each other; and the physically
relevant point, $\ell = t/\gamma$ ($\xi^2 = (1-v^2)$) is at a small
separation if $\phi$ is also close to 0 or $\pi$.  For all other cases
the equations predict smooth evolution with scale.

\begin{figure}
\epsfxsize=0.24\textwidth\epsfbox{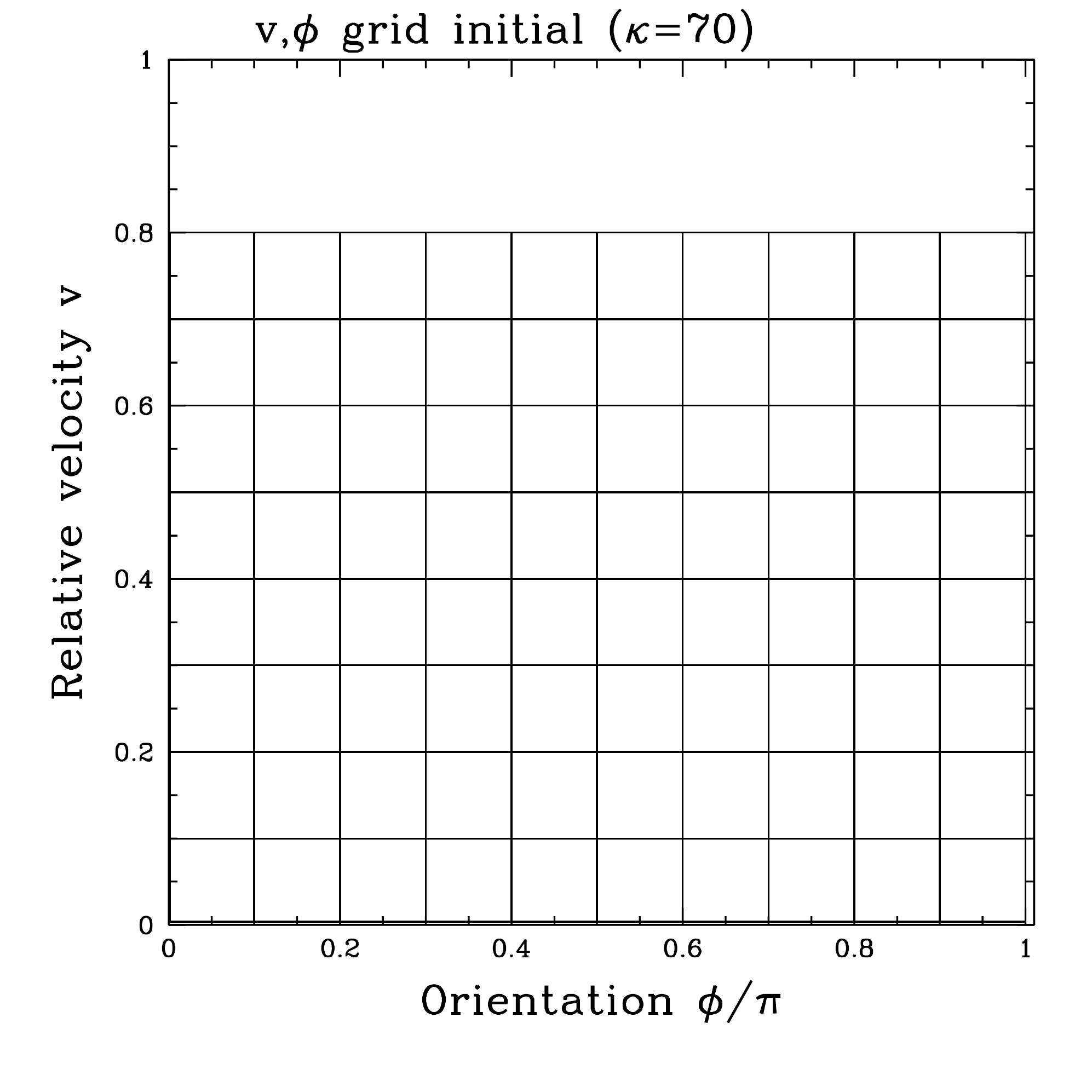} \hfill
\epsfxsize=0.24\textwidth\epsfbox{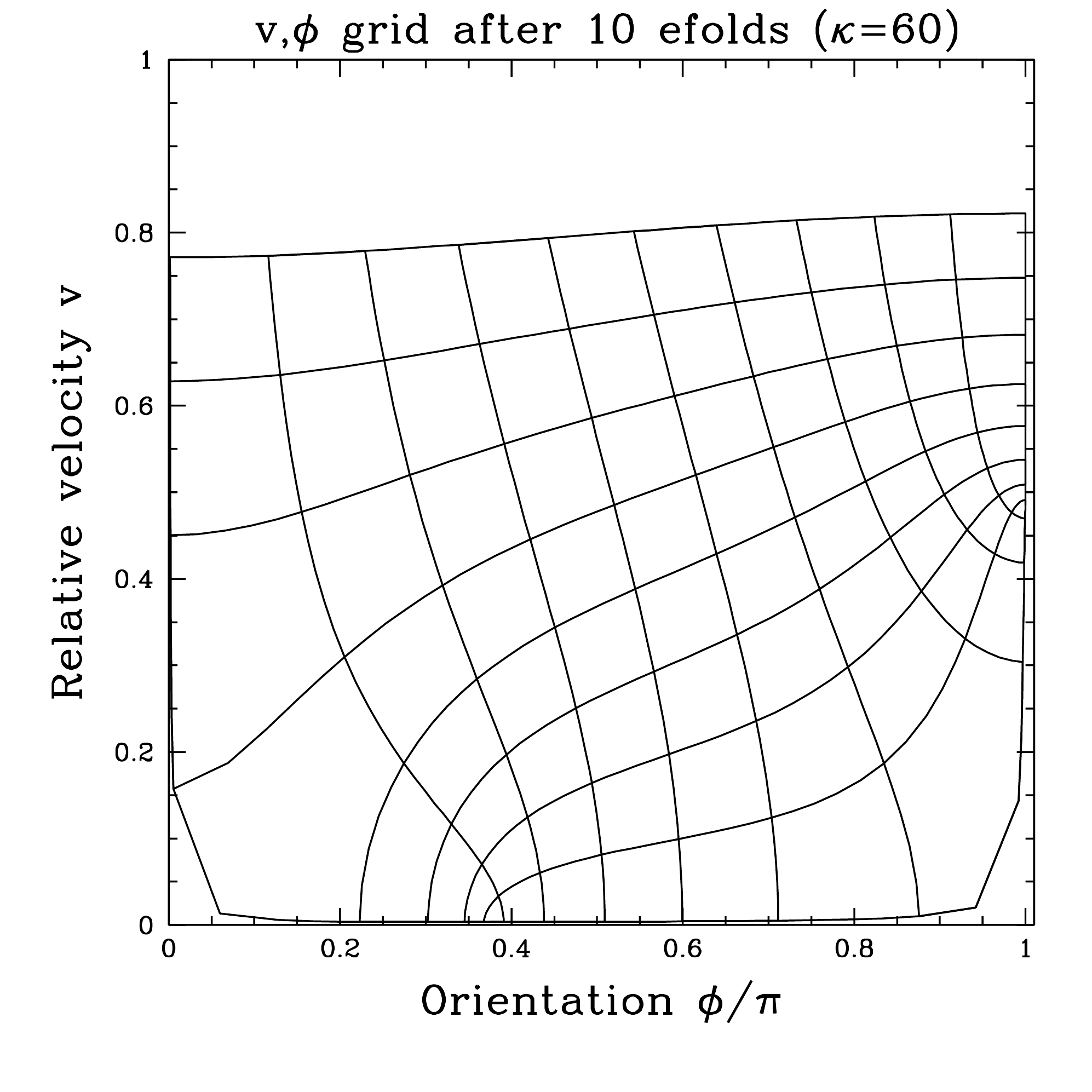} \hfill
\epsfxsize=0.24\textwidth\epsfbox{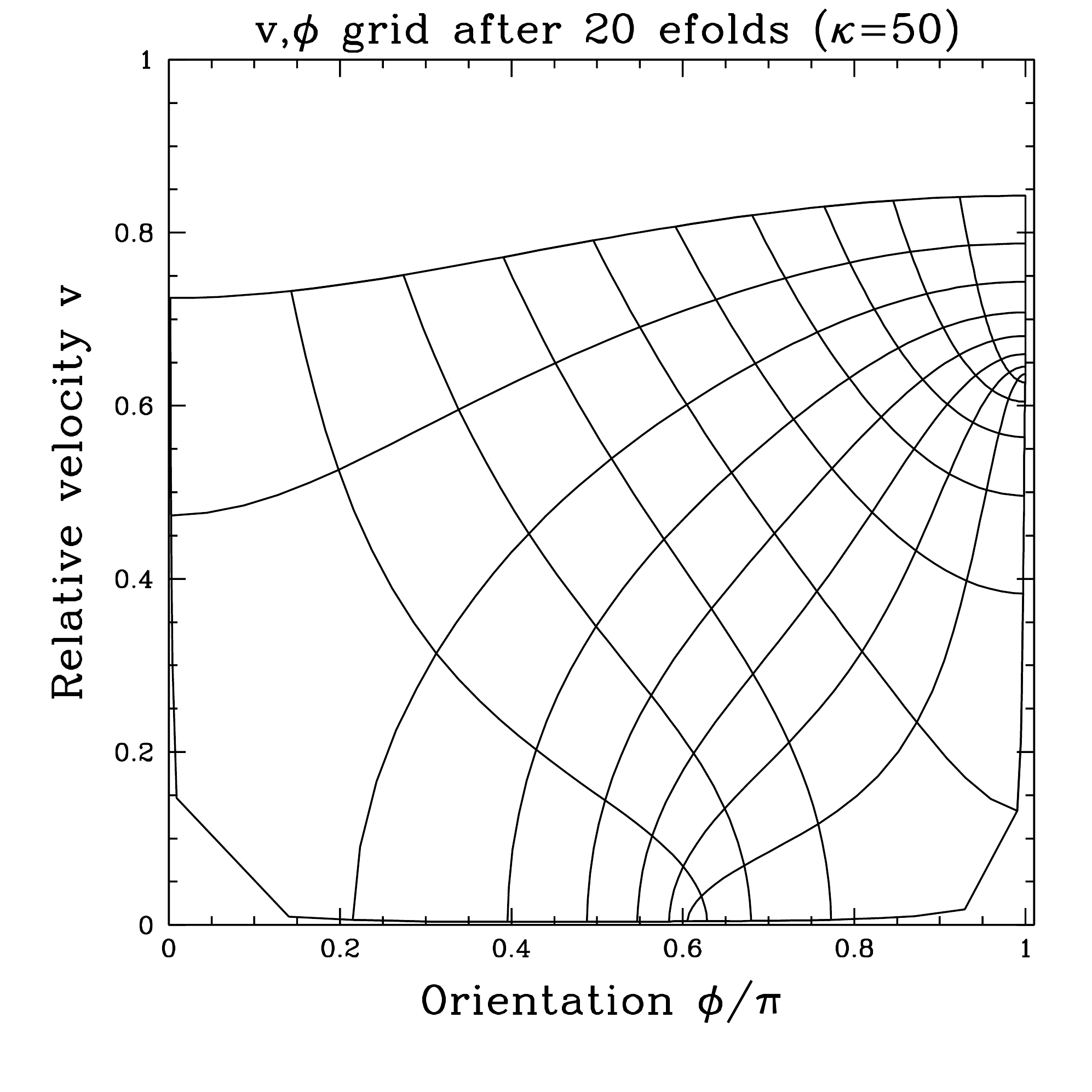} \hfill
\epsfxsize=0.24\textwidth\epsfbox{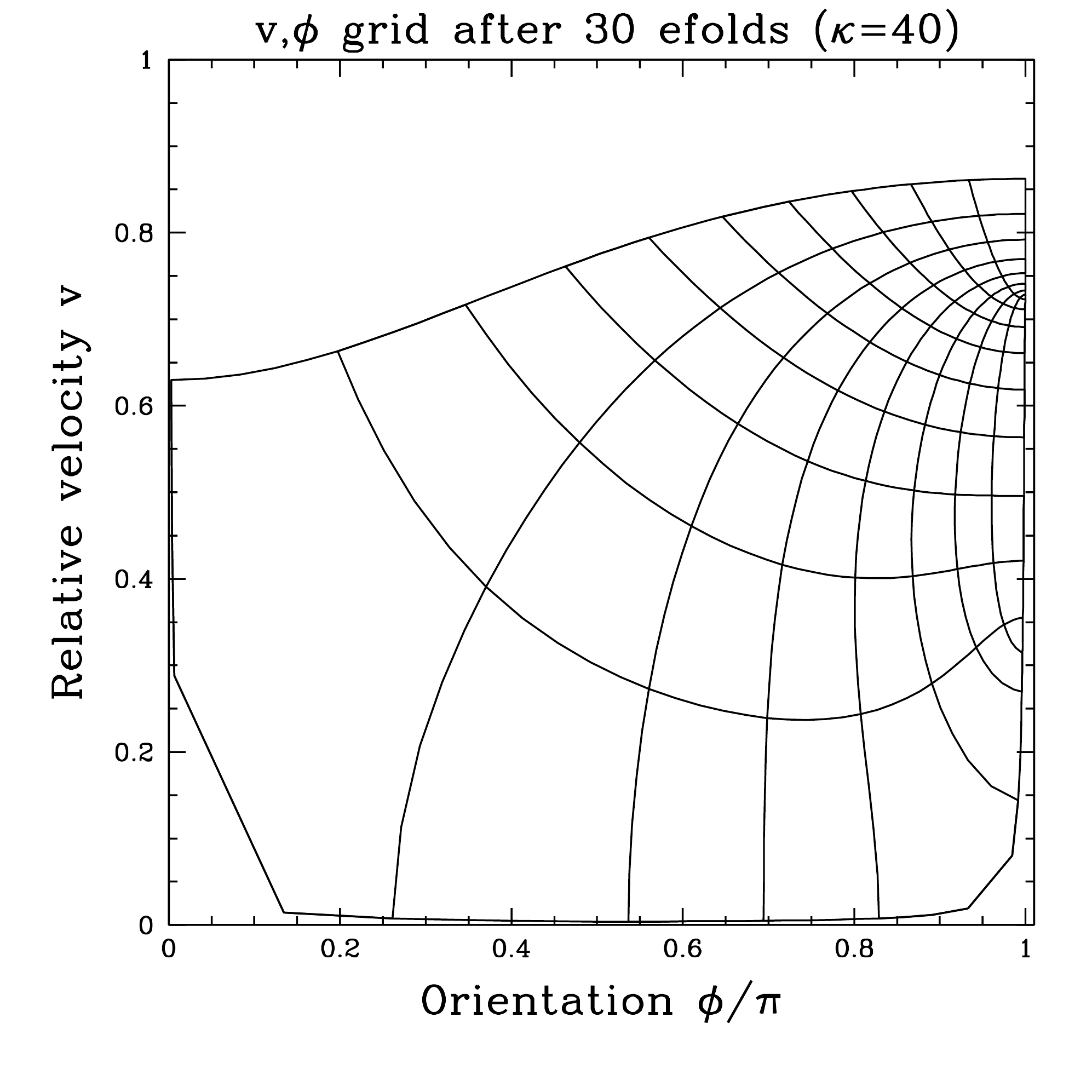} \hfill
\epsfxsize=0.24\textwidth\epsfbox{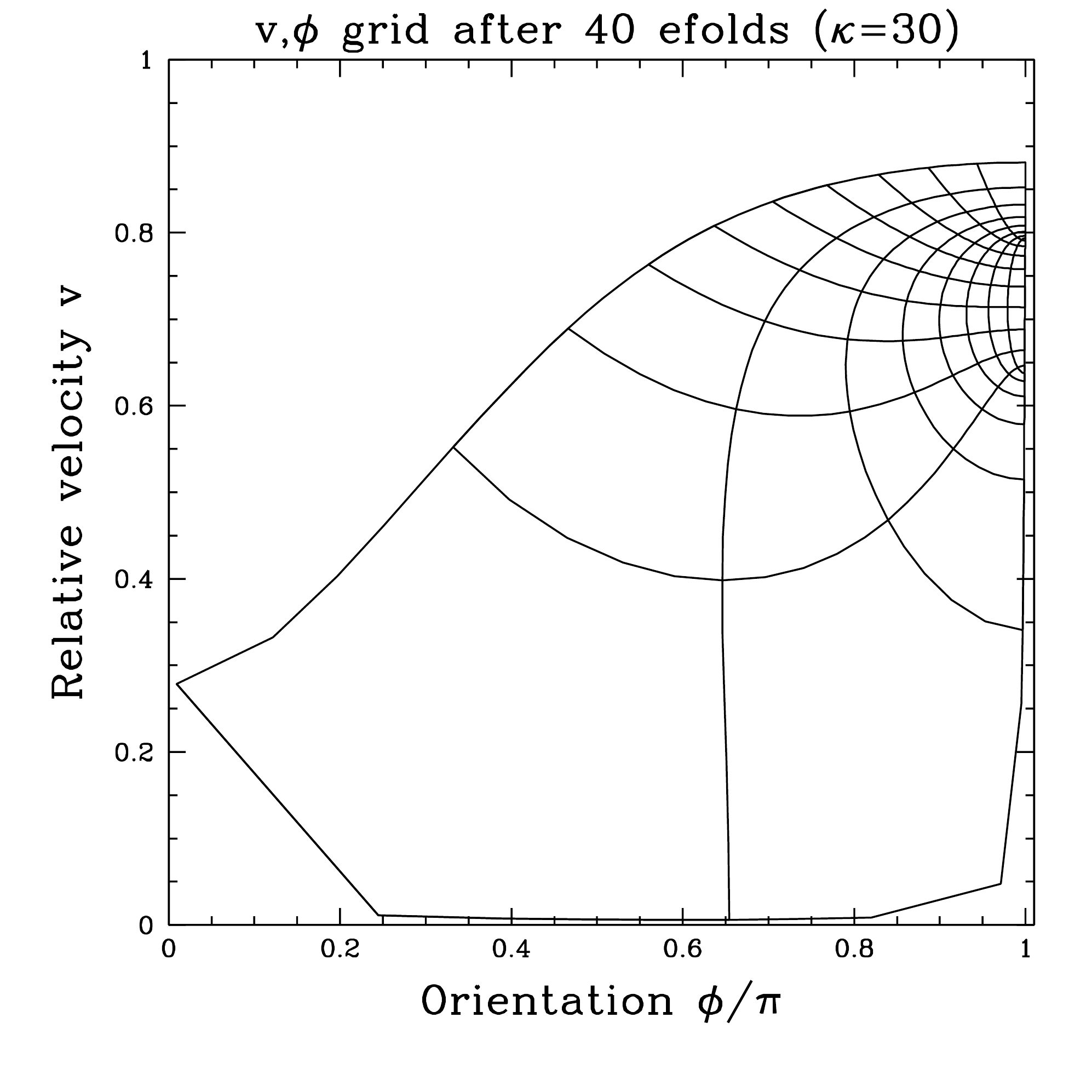} \hfill
\epsfxsize=0.24\textwidth\epsfbox{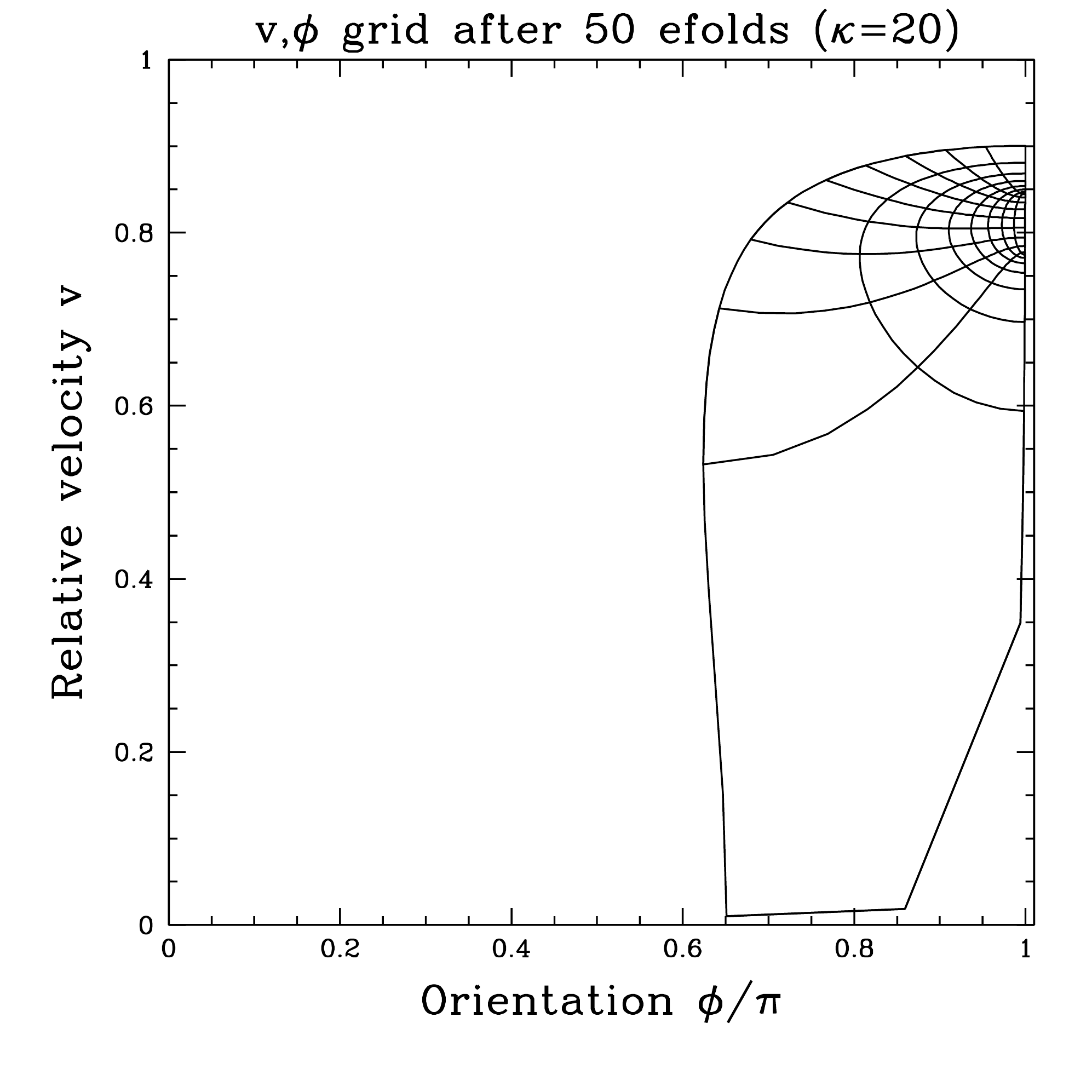} \hfill
\epsfxsize=0.24\textwidth\epsfbox{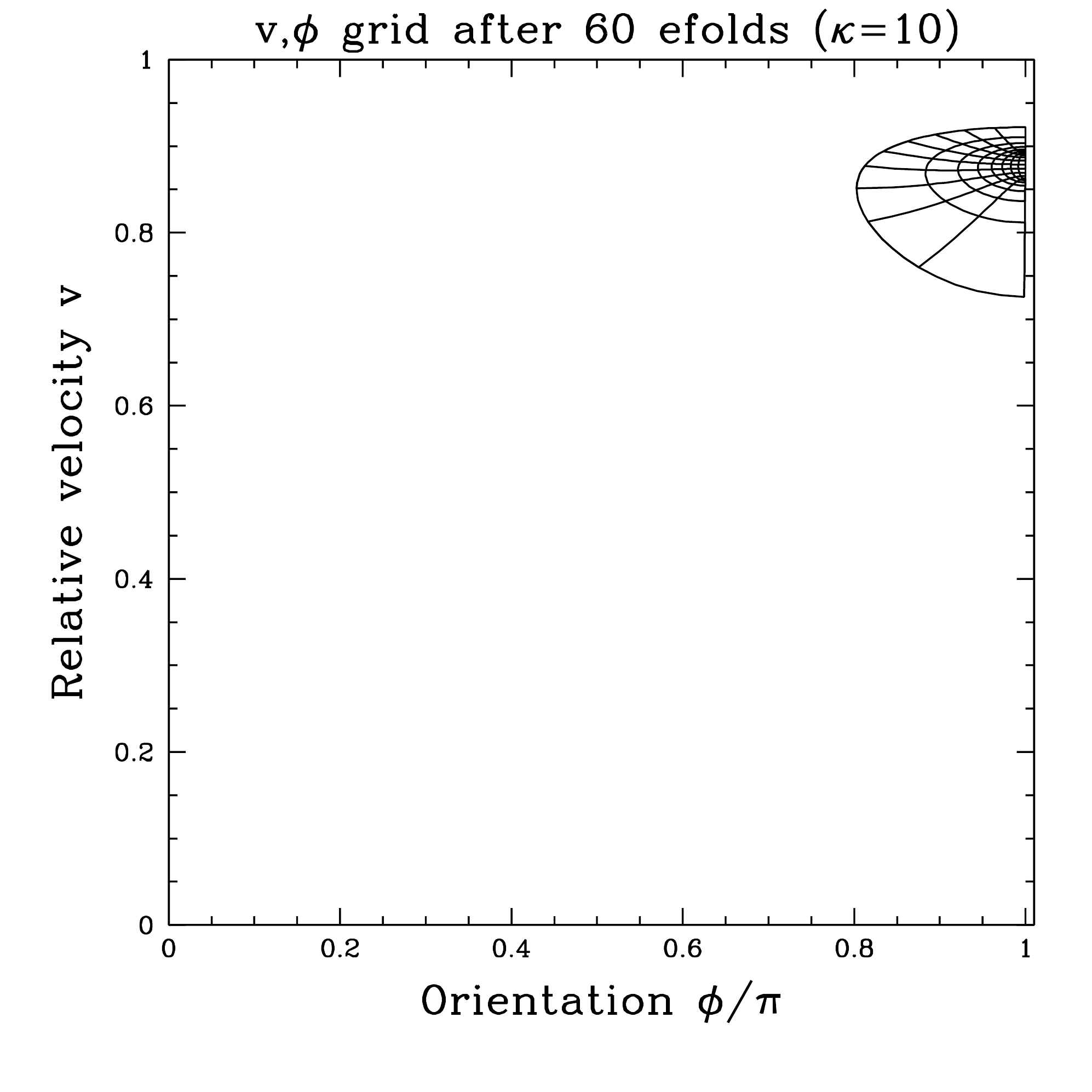} \hfill
\epsfxsize=0.24\textwidth\epsfbox{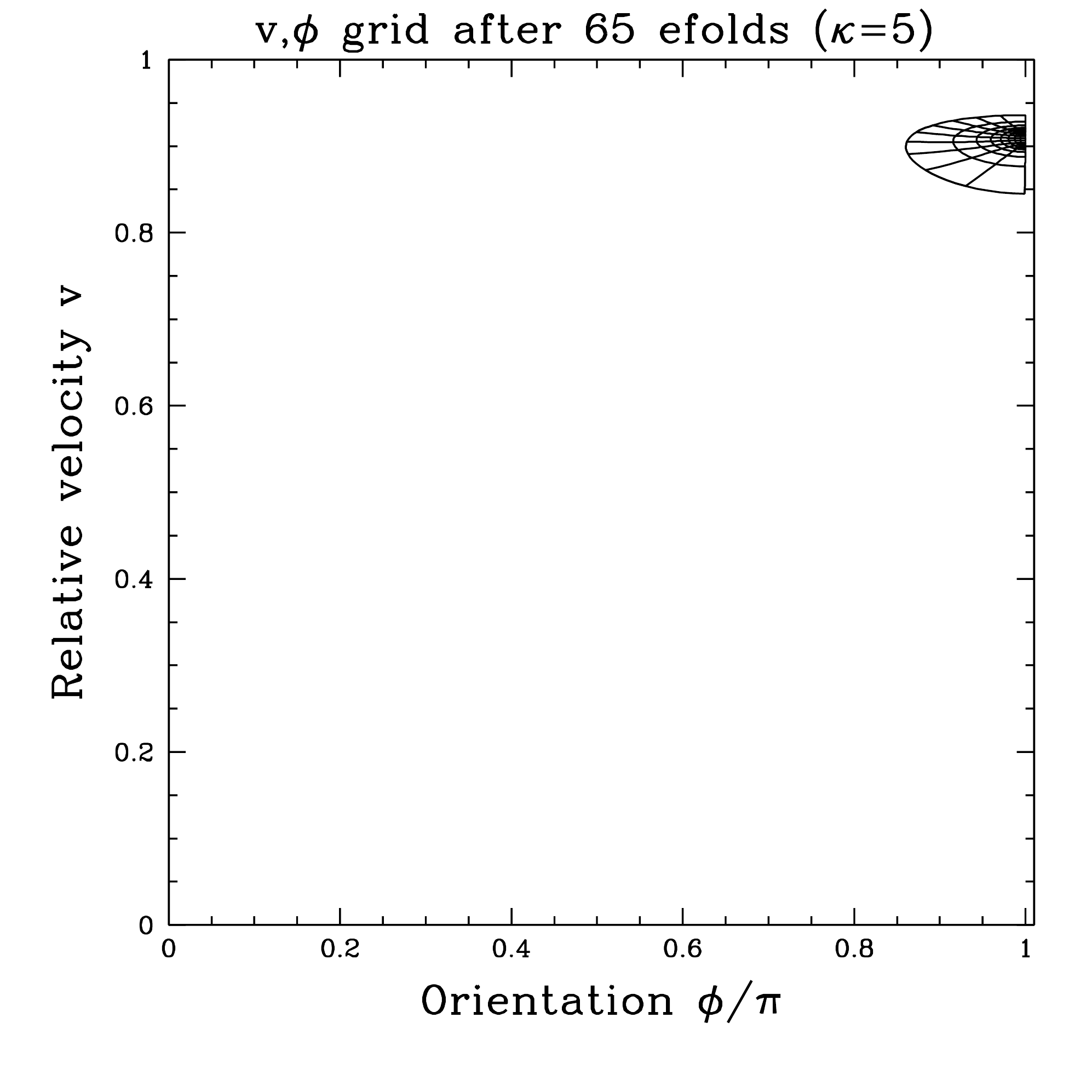} \hfill
  \caption{\label{fig:grid}
    Evolution of $\phi$ and $v$ ($x$ and $y$ axes) with logarithmic
    scale, $\kappa \equiv \ln(z m_h)$.  Starting with a grid over
    $(\phi,v)$ space at large separation ($\kappa=70$, our
    axion-motivated choice for an initial scale hierarchy) in the top
    left, we see how the initial $(\phi,v)$ value evolves in steps of
    10 in $\kappa$ or $e^{-10}$ in scale.}
\end{figure}

Figure \ref{fig:grid} shows how $(\phi,v)$ evolve with scale.  To see
the final value of $(\phi,v)$ for some initial choice, find the grid
position corresponding to the initial choice in the upper left frame
of the figure, and follow that grid location as the grid evolves
through the figure frames.  We have only considered
$v_{\mathrm{init}} < 0.8$, and we have taken the initial scale
hierarchy to be $\kappa = 70 = \ln(m_h/H)$, corresponding to
$m_h \sim 10^{11}$GeV and $H \sim T^2/m_{pl} \sim 10^{-19}$GeV as
motivated by an axionic string network at the scale where the network
breaks up.  We have followed the evolution to $\kappa = 5$, that is,
where the string separation is about 100 times larger than the string
core size, which is where numerical lattice studies can take over.
These choices are not essential; the main conclusion is that $\phi$
rotates to be very nearly antiparallel, and $v$ grows until the
$(1-v^2)^2$ term in \Eq{zmotion3} slows down its evolution.

\section{Microscopic study of intercommutation}
\label{sec:micro}

The most important result of the last section is that, for virtually
all initial string velocities and angles, the microscopic encounter
occurs at very high velocity and $\phi \simeq \pi$.  Specifically,
Figure \ref{fig:grid} shows that, for $v_{\mathrm{init}} < 0.6$, the
final velocity is in the range $v \in [0.887,0.924]$ and the angle is
in the range $(\pi-\phi) < 0.18$.  Therefore the only case of physical
relevance is nearly antiparallel strings approaching each other at
high velocity $v \simeq 0.9$.

Previous numerical studies \cite{Shellard:1987bv,perivolaropoulos:1991du}
have found that global strings approaching each other under these
conditions pass through each other without reconnection.  We will
revisit this conclusion and find that it arose from considering
insufficiently large boxes -- the parts of the string pair approaching
each other at impact parameter $m_h b \in [20,70]$ are essential to
see that interconnection does in fact occur.

In studying the string collision problem, we will use a simplifying
approximation, already considered by previous workers
\cite{Shellard:1987bv,perivolaropoulos:1991du}.  Consider again two
strings approaching each other; take the strings to approach along the
$z$ axis and to stretch primarily along the $y$ axis, with each
string's $x$ position determined by $x = \pm y \tan(\phi/2)$.  If the
strings are exactly parallel, then the
problem reduces to a 2-dimensional problem in which we ignore the $y$
coordinate.  If instead the strings are only very nearly
parallel, we can make a similar simplification, up to small
corrections.  The equation of motion for the scalar $\varphi$ field,
derived from \Eq{Lagrangian}, is
\begin{equation}
  \label{eq:EOM}
  \partial_t^2 \varphi = \partial_x^2 \varphi + \partial_z^2 \varphi
  + \partial_y^2 \varphi -V'(\varphi) \,.
\end{equation}
The field is expected to vary slowly in the $y$ direction, in which
case we can drop the $\partial_y^2 \varphi$ term here, with
corrections of order $(\tan \phi)^2 \ll 1$.  Then for each value of $y$, the
problem becomes a scattering problem of vortices in 2+1
dimensions, with impact parameter $b$ varying along the $y$-axis
as $b = 2x(y) = 2y \tan(\phi/2) \simeq y \phi$.
This approximation can break down at large times for some $y$ values.
Specifically, defining the location of the vortex in this 2+1D problem
to be $(x_v(b,t),z_v(b,t))$, then our approximation has broken down
if $dx_v/db > 1/\phi$ or $dz_v/db > 1/\phi$.  This will eventually
happen, but only at large enough $t$ that interconnection has already
happened.

We have studied this problem by implementing \Eq{eq:EOM} (without the
$\partial_y^2 \varphi$ term) on a 2+1D lattice.  We use a
next-nearest-neighbor improved gradient implementation, and a square
box which is $L \geq 400/m_h$ on a side.  Our initial conditions have
two strings, moving with velocities $v = \pm v_0 \hat{z}$ and located at
$(x,z) = ((L\pm b)/2,(2\mp 1)L/4)$.  That is, they are separated by
the impact parameter $b$ in the $x$ direction and by $L/2$ in the $z$
direction. Our starting condition is that $\varphi/f_a$ is the
complex-number product of the solution of each moving string, and our
boundary condition is that we continuously enforce this condition
within 2 lattice units of the boundary.  This choice avoids any unphysical
forces from mirror-image charges, but it means that our simulation can not
be trusted and must be stopped after information about the real
trajectory of a string has had time to propagate to the boundary and
back to the string, roughly $t = L/2$ time.  This is why we start the
strings some distance off the boundary.  For $v=0.9$ we use a
lattice spacing of $m_h a=0.5$, confirming on $m_h a=0.33$ lattices
that the results are not sensitive to the spacing.  We also study
$v=0.95$ and $v=0.975$ with $m_h a = 0.25$ and $m_h a = 0.125$
lattices respectively.

\begin{figure}
  \centerline{\epsfxsize=0.6\textwidth\epsfbox{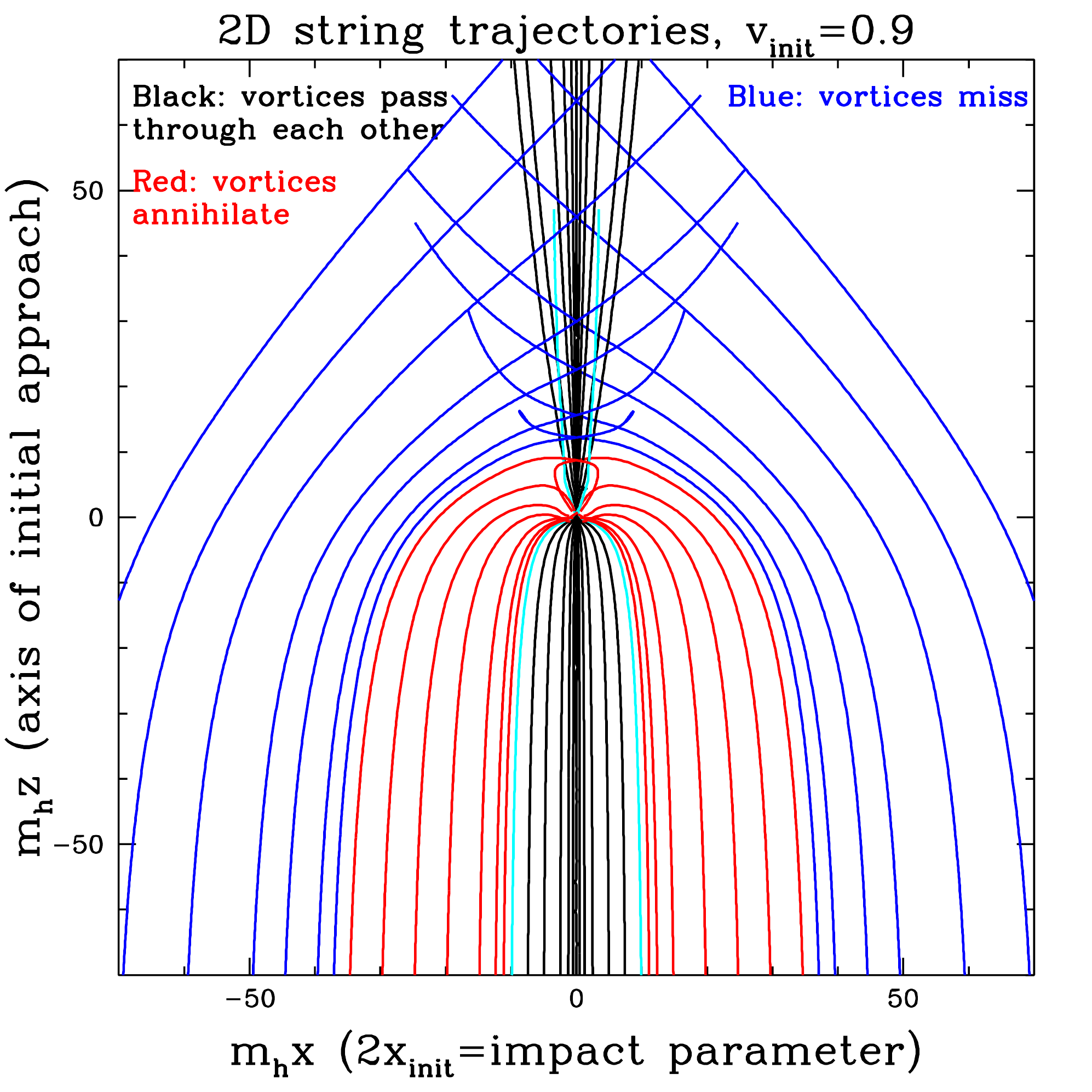}}
  \caption{\label{fig:traject}
    Each line represents the trajectory of a vortex, approaching an
    antivortex at $v=0.9$; each trajectory starts with a different
    impact parameter.  For small impact parameters (black lines) the
    vortex passes through the antivortex and keeps moving.  For
    intermediate values (red lines) the vortex and antivortex attract
    each other, collide, and annihilate.  For large impact parameters
    (blue), they miss and keep moving, with a reduced velocity.
  }
\end{figure}

We find, in agreement with previous studies, that a vortex-antivortex
pair approaching each other with $v=0.9$ at initial separation
$\Delta z = 200/m_h$ and zero impact parameter pass straight through
each other with only a small reduction in velocity.  But
for the strings as a whole to pass through each other, strings must
survive passing each other at \textsl{every} impact parameter.
Instead, we find for $v=0.9$ that there is a range of impact
parameters, $m_h b \in [21,70]$, over which the vortex and antivortex
curve towards each other and annihilate.  For larger $b$ the vortex
and antivortex bend as they swing past each other and radiate
significant energy, losing much of their velocity; but they escape.
We illustrate this behavior in Figure \ref{fig:traject}
by plotting the trajectory of the upwards-moving vortex (the
lower-moving antivortex is an inversion image through the central
point) for a number of initial impact parameters.  Each curve in the
figure represents the trajectory (time-history) of a vortex, entering
the plotted region from the bottom as an antivortex enters from the
top.  For small impact parameter (black lines), the vortices pass
through each other at the origin and continue.  The cyan trajectory is
the last with this property; it slows to a stop at the end of the
simulation and will fall back down and annihilate if we follow the
simulation longer.

For the next-larger range of impact parameters (red lines), the
strings curve to hit each other and annihilate.
For still larger impact parameters (blue lines), the trajectories
curve but miss.  In this
range, the smaller the impact parameter the more energy is radiated
and the slower the string emerges.  These strings will also eventually
attract each other and collide, but not in the time range available in
our study.

\begin{figure}
\epsfxsize=0.24\textwidth\epsfbox{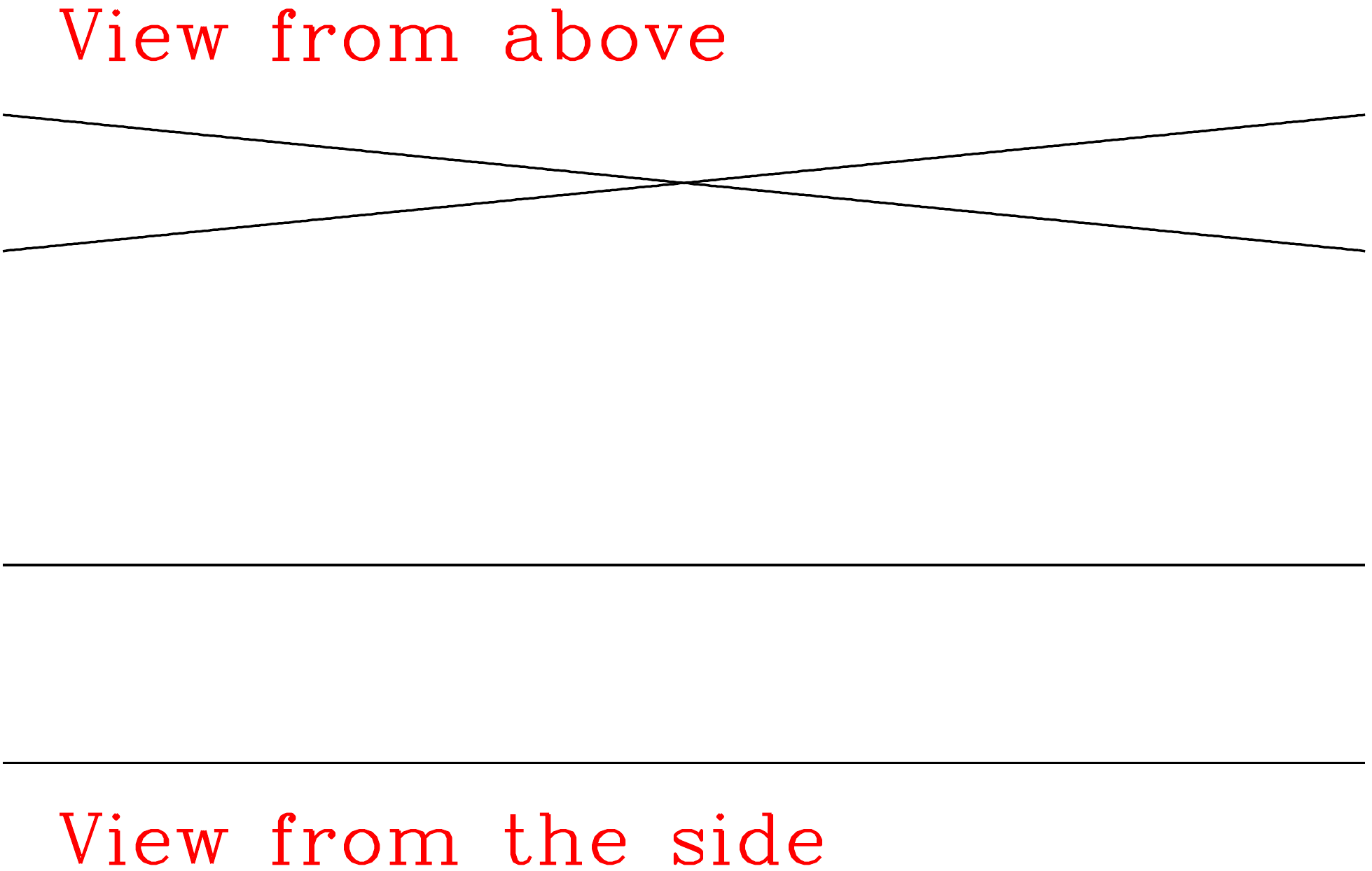} \hfill
\epsfxsize=0.24\textwidth\epsfbox{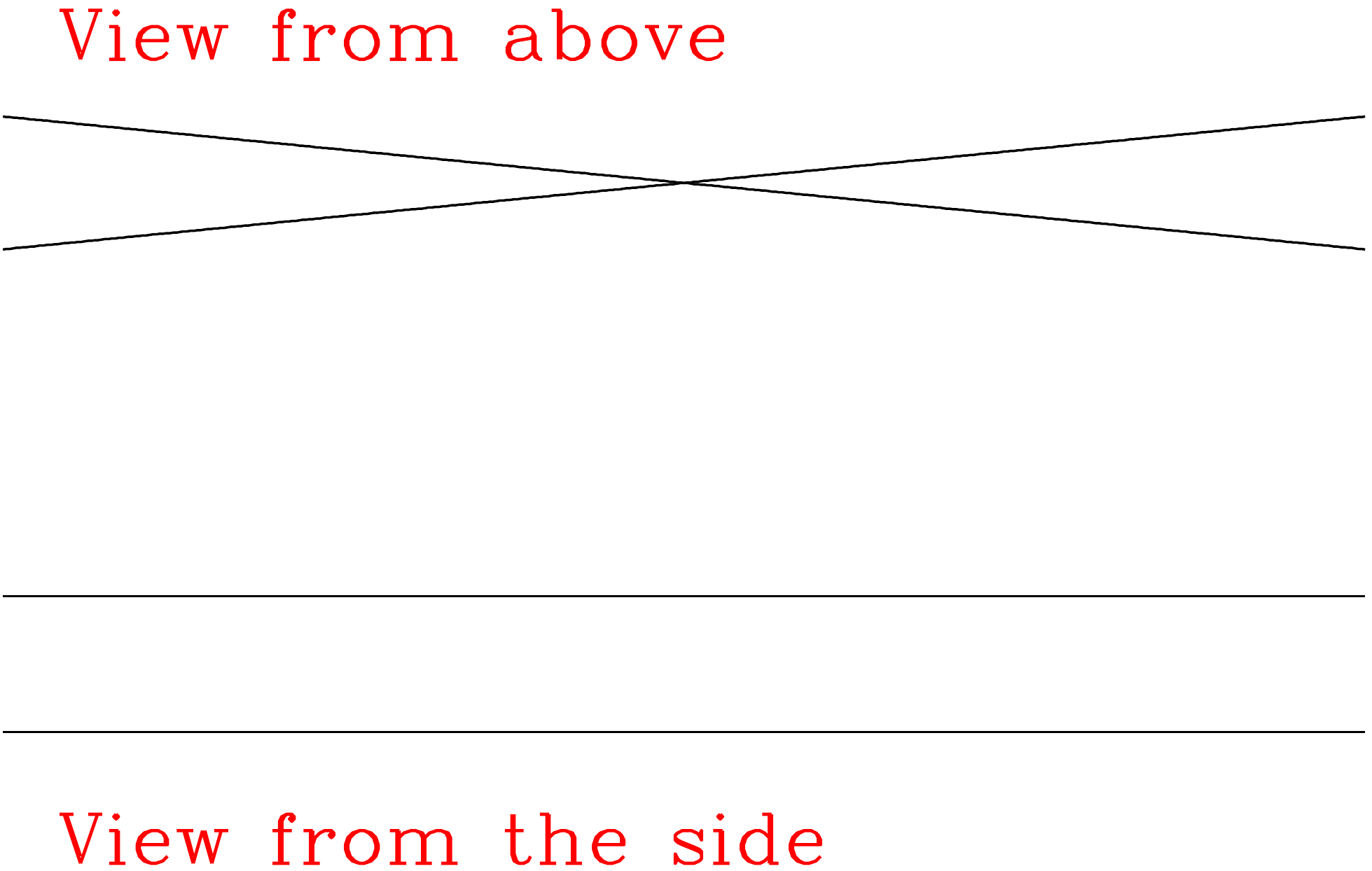} \hfill
\epsfxsize=0.24\textwidth\epsfbox{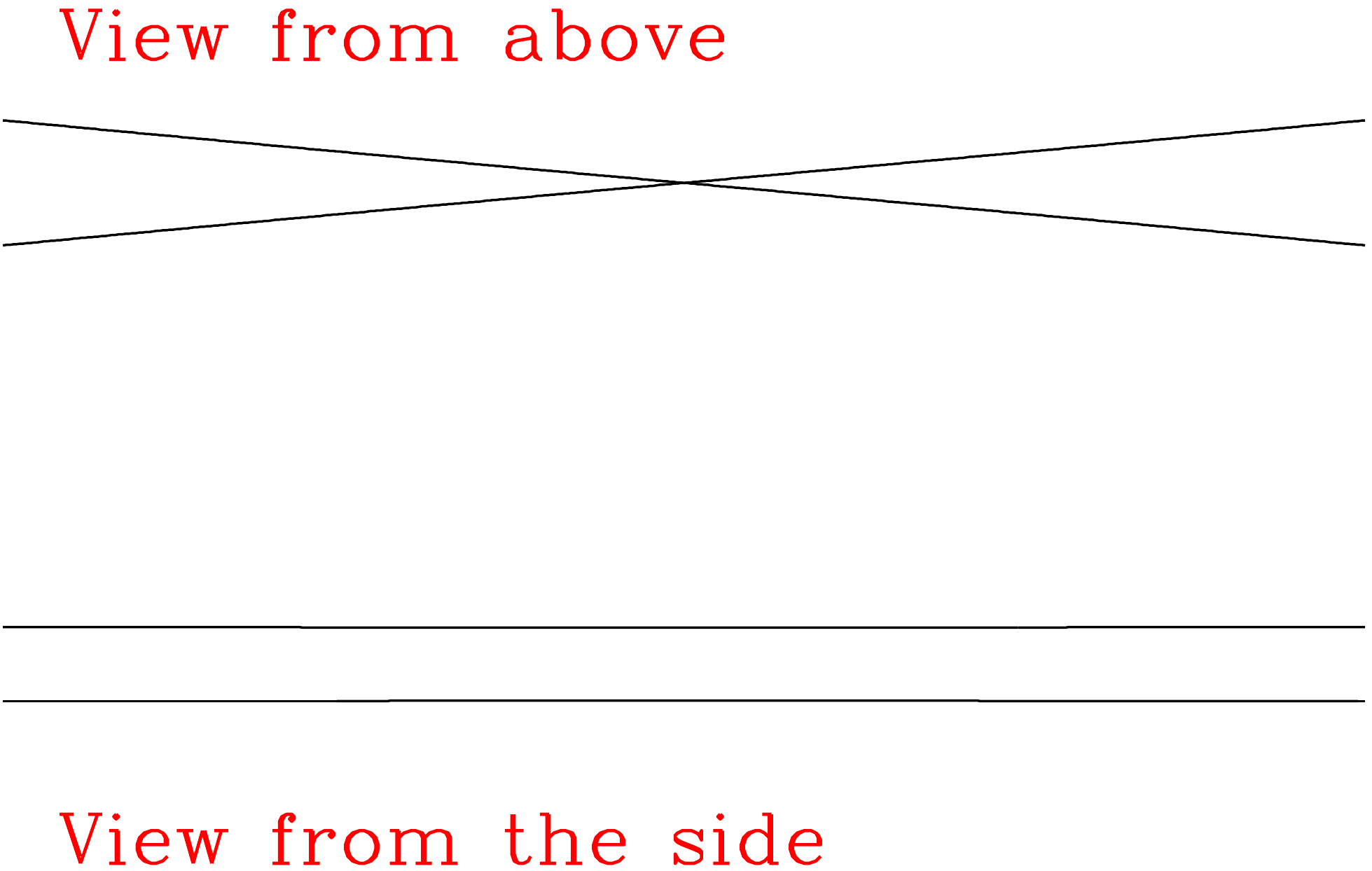} \hfill
\epsfxsize=0.24\textwidth\epsfbox{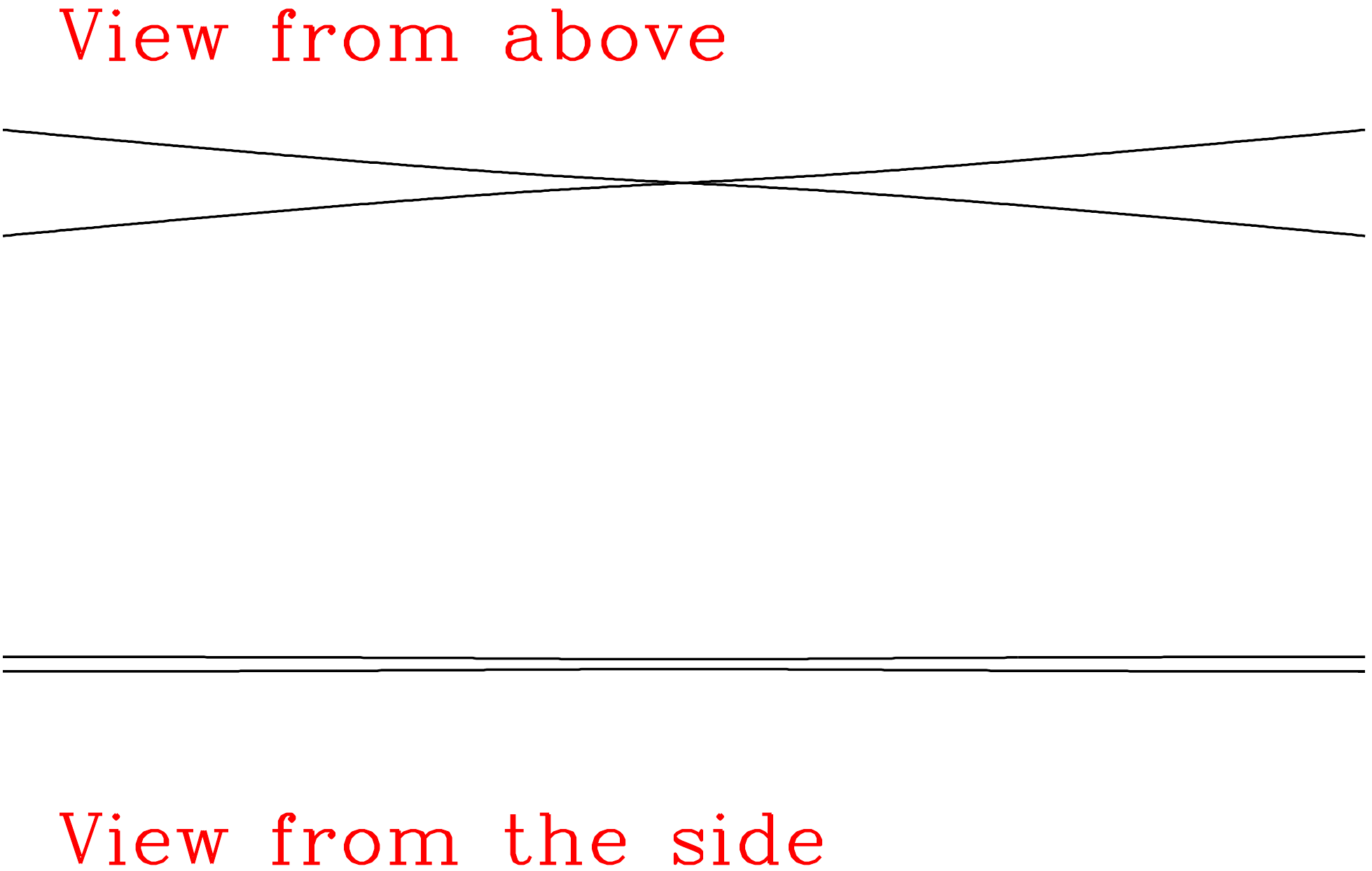} \hfill

\vspace{1.9ex}

\hrule

\vspace{1.9ex}

\epsfxsize=0.24\textwidth\epsfbox{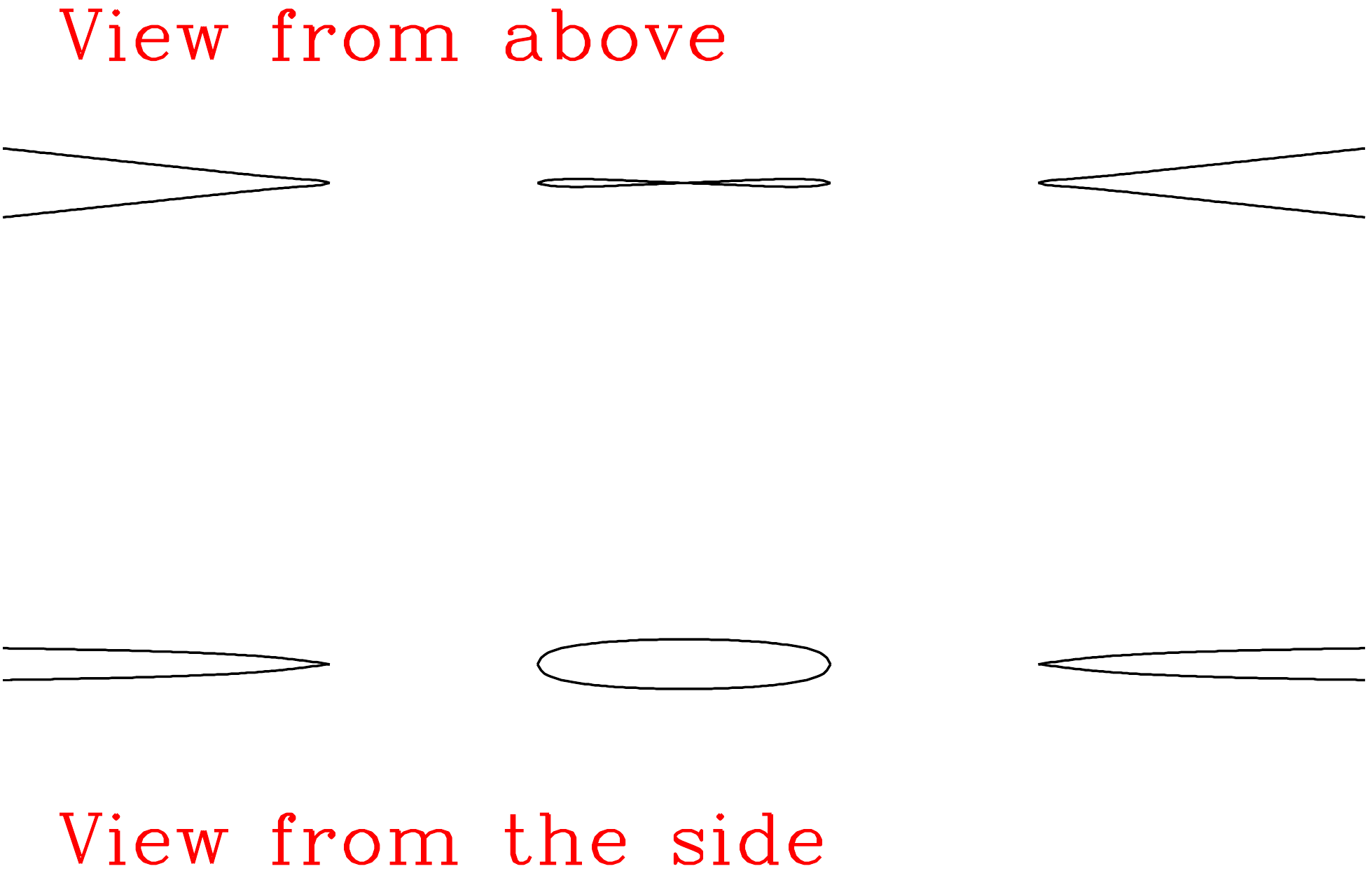} \hfill
\epsfxsize=0.24\textwidth\epsfbox{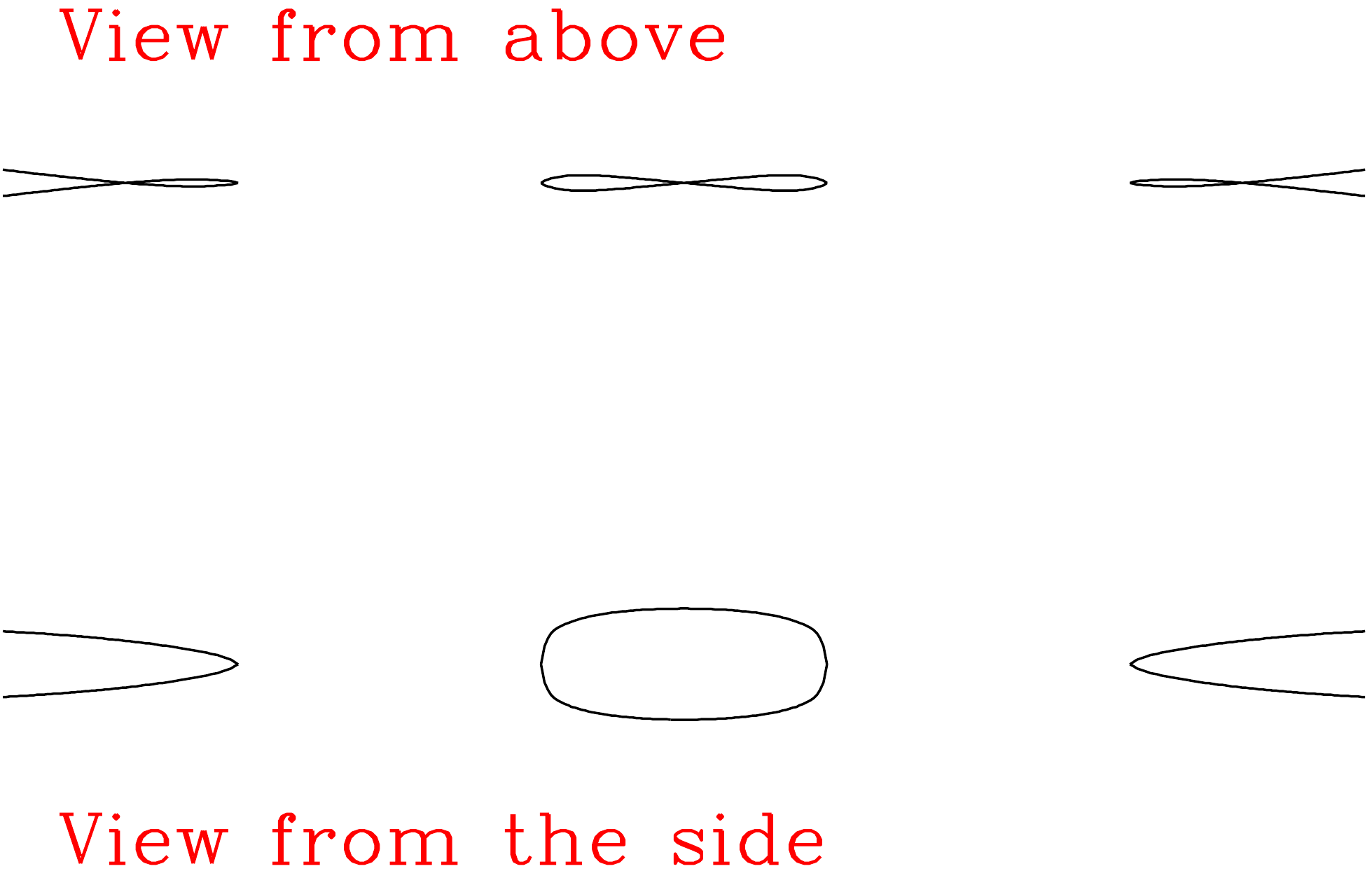} \hfill
\epsfxsize=0.24\textwidth\epsfbox{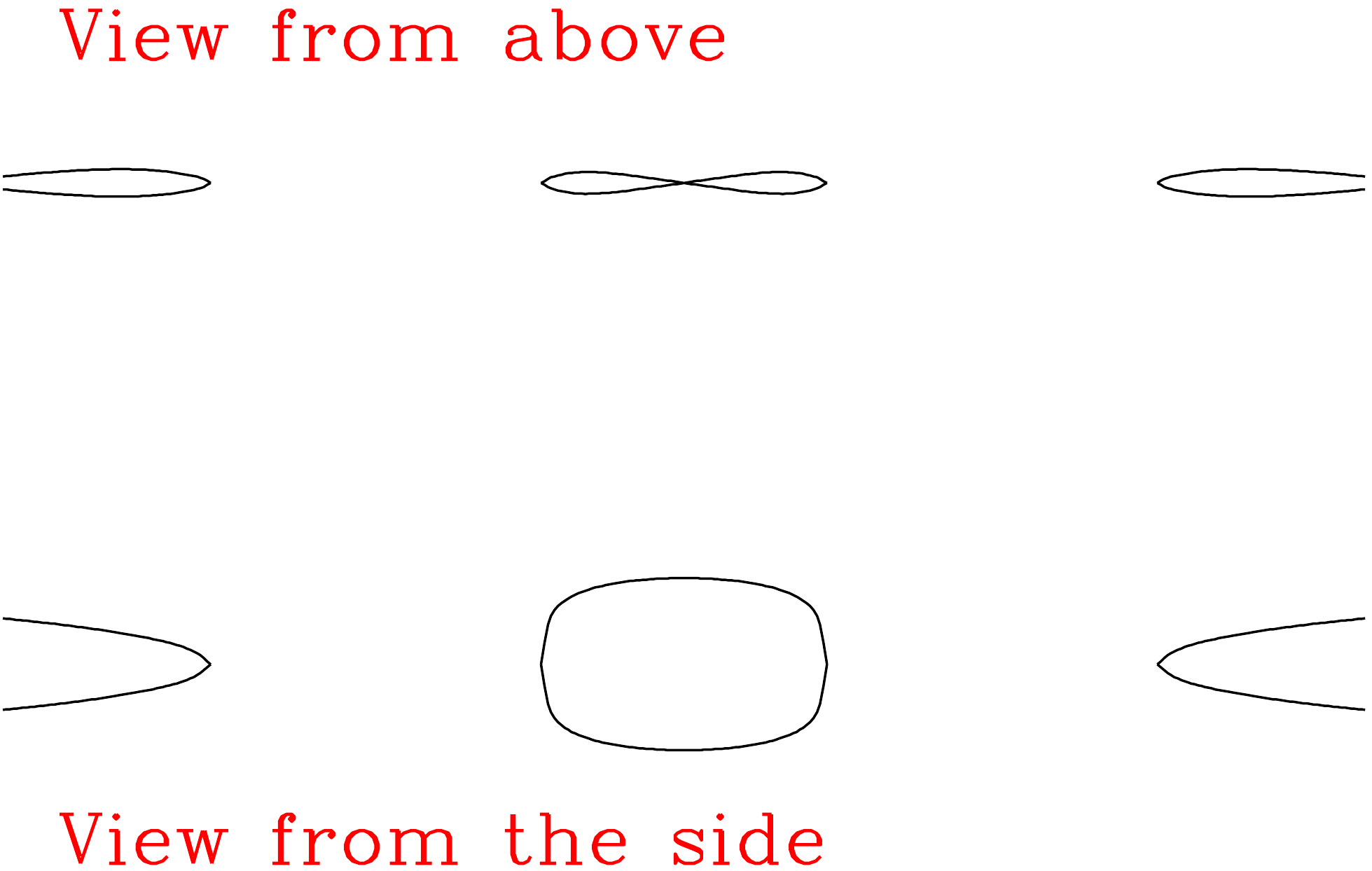} \hfill
\epsfxsize=0.24\textwidth\epsfbox{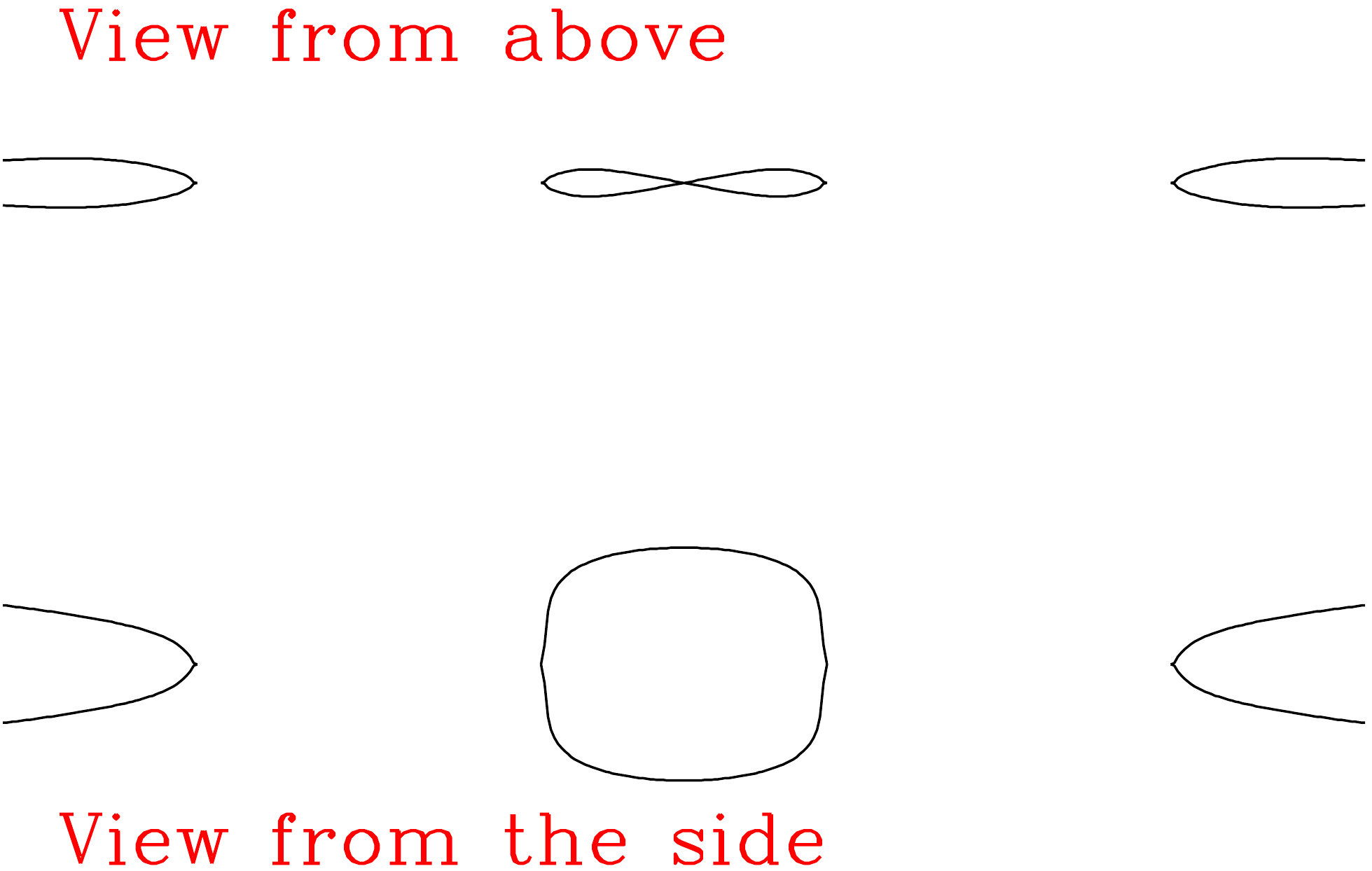} \hfill
  \caption{\label{fig:frames}
    Several single-time frames in a movie of string collision at large
    velocity and nearly antiparallel orientation, made by solving
    the 2+1D problem independently in each $(x,z)$ plane.  In the
    first 4 frames, the strings move towards each other.  In the fifth
    frame (leftmost on the second row), they have just crossed. While the
    strings pass through each other at the center, there is a region
    to either side where they annihilate, ensuring intercommutation.}
\end{figure}

We can sew these trajectories together into a string collision picture
using the prescription described around \Eq{eq:EOM}.  We solve
the 2+1D vortex-scattering problem for a range of impact parameters,
and we compile the solution for each impact parameter at one fixed
time to give a single-time ``frame'' in a movie of the interaction of
a pair of strings.  We show several equal-time ``movie frames'' in
Figure \ref{fig:frames}.  In other words, we determine the string
locations in the $(x,z)$ plane at time $t$ and at a given $y$ by
evolving a vortex-antivortex pair with impact parameter
$b=2y\tan(\phi/2)$ for a time $t$, and then connect together these
points at different $y$ to find the string.  In making the figure we
chose $\phi=1/5$; making a different choice $\phi_{\mathrm{new}}$
corresponds to stretching the long axis by a factor of
$\phi_{\mathrm{old}}/\phi_{\mathrm{new}}$.  At or near the point where
the strings meet, they pass through each other.  On either side of
this region there is a region where they annihilate.  Far from the
intersection point, the strings miss each other.  The result is a loop
of string at the center, separated by gaps, with the ends reconnected.
We include the full movie of the collision in the extra
materials accompanying this paper.

In the last three frames of Figure \ref{fig:frames}, the string loop
in the middle has portions where the assumption of smooth
$y$-variation is violated.  Therefore the evolution of this loop is
not well described.  This occurs because the vortex evolution is very
sensitive to the impact parameter right at the critical value where
the strings either pass through each other or annihilate (around
$b=20/m_h$).  But the region where the strings annihilate is wide
enough that this cannot causally affect the fact that the central loop
is separated from the string on the left and right.  Therefore the
result that the strings annihilate on either side of this loop is
robust.

We repeated the analysis leading to Figure \ref{fig:traject} for
$v=0.80$, for $v=0.95$, and for $v=0.975$ (using $am_h=0.25$ for
$v=0.95$ and $am_h = 0.125$ for $v=0.975$),
and found that the same qualitative behavior occurs, but
with a slightly different range of impact parameters where string
annihilation occurs.  For $v=0.8$ it occurred in $m_h b \in [24,93]$,
for $v=0.95$ it occurred in $m_h b \in [19,62]$, and for
$v=0.975$ it occurred in $m_h b \in [20.5,56]$.  Our criterion was
that the strings had annihilated and were absent at a time
$t = 100/m_h$ after the expected intersection time.
Therefore, strings colliding at a (microscopically) nearly antiparallel
relative angle will intercommute for all velocities $v < 0.975$ (

\section{Discussion and Conclusions}
\label{sec:conclusion}

We have shown that the long-range interactions between global strings
play an important role when two strings approach to cross each other.
The inter-string forces, operating over a huge logarithmic range of
scales, ensure that the microscopic crossing occurs at very high
velocity and very nearly antiparallel approach.  We also demonstrated,
contrary to previous studies, that such high-speed nearly antiparallel
string collisions result in intercommutation.  Our investigation was
based on the fact that, for nearly parallel or antiparallel approach, we can
solve the behavior in 2+1 dimensional slices, and that when a 2+1D
vortex antivortex pair approach each other relativistically, there is
a wide range of nonzero impact parameters over which the strings bend to
collide and annihilate.  For small impact parameters they pass through
each other, but that only causes a small isolated string loop at the
string-crossing location; it does not prevent intercommutation.

This result looks peculiar, and it is worth pausing a moment to see if
we can understand it.  A key feature of those trajectories in Figure
\ref{fig:traject} which annihilate (the red trajectories in the
figure) is that they all have the string's trajectory bend by a large
angle before the annihilation.  How large a bending angle do we expect?
In the approximation that the string's
trajectory does not deflect, we find by integrating the second line of
\Eq{Fresult} over time $\int dt = \int dz/2v$ that the transverse
momentum absorbed as the vortices pass each other is
$2\pi^2 f_a^2 (1+v^2)/2v \simeq 2\pi^2 f_a^2$.  This is to be compared
to the total momentum of the vortex, $\gamma v \pi f_a^2 \ln(b m_h)$.
For $\gamma \sim 2$ and moderate impact parameters $b$, the string is
expected to bend by a large angle.
A rapid, large deflection by a very relativistic charge results in a
large radiated power.  The vortex thereby loses a large fraction of
its energy.  The Coulomb potential is confining in 2+1 dimensions, and
the vortex-antivortex pair become tightly bound and fall onto each
other at low velocity.  And a low-velocity collision does cause
annihilation.

This should be the behavior in a range of impact parameters.  For
larger impact parameter, $\ln(b m_h)$ is larger and the string has
more inertia to absorb the bending force.  In this case the energy
radiated is smaller and the strings do not become tightly bound;
instead the string escapes (at least for long time scales).  On the
other hand, for very small impact parameter, a small bend is enough
for the vortices to meet head-on.  Little energy is radiated before
the collision, and so the strings meet at high velocity and have
enough energy to pass through each other.

Our study shows that strings should intercommute if their macroscopic
relative velocity is below $v=0.8$.  We expect this to cover the vast
majority of string collisions in a network evolution (recall that this
is the relative velocity; to achieve it each string must move at
$v>0.8$ straight at the other string).  Nevertheless, collisions
may occasionally occur with larger velocity.  In this case it is no
longer true that the microscopic collision angle is small, see
Figure \ref{fig:grid}.  Our analysis is not valid here, and
it is not clear to us what happens in this case.  But previous work
\cite{Shellard:1987bv} suggests that intercommutation should also
occur in this case.

Also note that we have only considered one model with global strings,
the complex scalar or O(2) (or relativistic $xy$) model.  Models with
more complicated vacuum manifolds may show different behavior and
would require a separate analysis.  This will be needed if any such
model proves to be of sufficient physical interest.

\section*{Acknowledgments}

We thank the Technische Universit\"at Darmstadt and its Institut f\"ur
Kernphysik, where this work was conducted.

\bibliographystyle{unsrt}
\bibliography{refs}

\end{document}